\documentclass[prb,aps,twocolumn,superscriptaddress]{revtex4-2}

\usepackage{amsmath,amssymb,amsfonts,bbm,color,graphicx}

\usepackage[caption=false]{subfig}

\usepackage[utf8]{inputenc}
\usepackage[T1]{fontenc}

\usepackage[unicode=true,colorlinks=true]{hyperref}
\hypersetup{linkcolor=blue,citecolor=blue,urlcolor=blue}

\usepackage{bm}
\usepackage{dsfont}
\usepackage[usenames,dvipsnames]{xcolor}
\usepackage{wasysym}
\usepackage{enumitem}

\usepackage{braket}

\newcommand{\iu}{\mathrm{i}} 
\newcommand{\eu}{\mathrm{e}} 
\newcommand{\du}{\mathrm{d}} 
\newcommand{\hc}{\mathrm{h.c.}} 

\newcommand{\bvec}[1]{\bm{#1}}

\newcommand{\Ztwo}{\mathbb{Z}_2}
\newcommand{\Uone}{\mathrm{U(1)}}
\newcommand{\SUtwo}{\mathrm{SU(2)}}

\begin{document}

\title{Mott insulators in moiré transition metal dichalcogenides at fractional fillings: Slave-rotor mean-field theory}

\author{Zhenhao Song}
\affiliation{Department of Physics, University of California, Santa Barbara, CA 93106, USA}
\author{Urban F. P. Seifert}
\affiliation{Kavli Institute for Theoretical Physics, University of California, Santa Barbara, CA 93106, USA}
\author{Zhu-Xi Luo}
\affiliation{Department of Physics, Harvard University, Cambridge, MA 02138, USA}
\author{Leon Balents}
\affiliation{Kavli Institute for Theoretical Physics, University of California, Santa Barbara, CA 93106, USA}
\affiliation{Canadian Institute for Advanced Research, Toronto, Ontario, Canada}

\begin{abstract}

In this work, we study a slave-rotor mean-field theory of an extended Hubbard model, applicable to transition metal dichalcogenide moir\'e systems, that captures both the formation of Wigner crystals as well as exotic spin states on top of these charge backgrounds.
Phase diagrams are mapped out for different choices of long-range Coulomb repulsion strength, reproducing several experimentally found Wigner crystal states.
Assuming unbroken time reversal symmetry, we find several spin liquid states as well as dimer states at fractional fillings.
While spin dimer states are always found to have the lowest mean field energy, several spin liquid states are energetically competitive and may be stabilized by including gauge fluctuations or further interaction terms.
We further discuss possible experimental signatures of these states pertinent to two-dimensional moiré heterostructures.

\end{abstract}

\date{\today}

\maketitle

\section{Introduction}
\begin{figure}[t]
    \centering
    \includegraphics[width=.8\columnwidth]{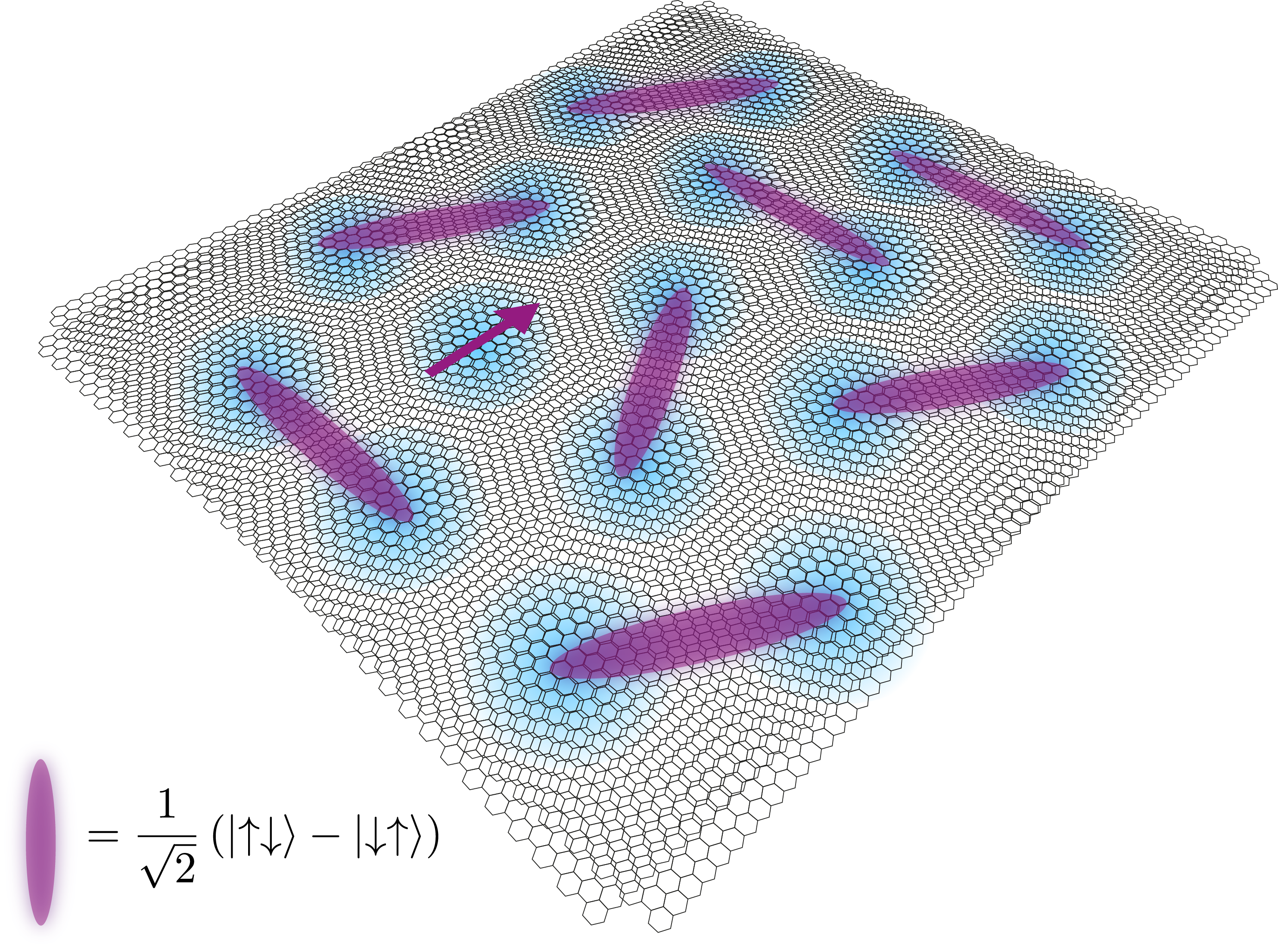}
    \caption{Illustration of quantum magnetism in moiré transition metal dichalcogenides. At low energies, charged quasiparticles carry a pseudospin-1/2 degree of freedom. In incompressible phases (for example at half-filling, pictured), the charge degrees of freedom (blue) become localized in Wannier orbitals of the moiré triangular lattice.  The remaining spin degrees of freedom are frustrated, and quantum fluctuations can stabilize quantum spin liquid states (for example resonating valence bonds (RVB), illustrated with valence bonds in purple, supporting isolated spinons as fractionalized excitations) or non-magnetic dimer states.}
    \label{fig:enter-label}
\end{figure}

Moir{\'e} patterns are formed by two or more similar two dimensional lattices overlaid with a slight relative strain or twist angle, giving rise to a large-scale periodic structure.
The most prominent example, twisted bilayer graphene, theoretically proposed by Bistritzer and MacDonald \cite{MacDonald11}, has been found to host a variety of novel phenomena such as superconductivity and correlated insulating states~\cite{cao18a,cao18b}. 
Moiré systems often feature strongly quenched kinetic energy scales and flat bands such that electronic interactions become dominant, providing a fertile ground for exploring strong correlations in condensed matter systems.
However, the flat band of twisted bilayer graphene is highly degenerate (spin and valley) and nonlocal \cite{song22}, such that the validity of Hubbard-type models for localized orbitals is under debate.

On the other hand, moir{\'e} heterostructures constructed from transition metal dichalcogenides (TMDs) also exhibit flat bands, where only a two-fold pseudospin degeneracy is present due to large spin-orbit coupling and resulting spin-valley locking in TMDs.
Wannier centers constructed from the flat band of some moir{\'e} TMDs turn out to be localized at triangular superlattice sites, and thus effective moir{\'e} Hubbard model can be formulated \cite{pan20a,pan20b}.
Moir{\'e} TMDs are also insensitive to the precise magic angle, i.e. flatness is maintained for a large range of twist angles.
These features, along with tunable filling via gating, make moir{\'e} TMDs an ideal and flexible platform to ``simulate'' \cite{tang20,kennes21} the Hubbard model and study novel phases that can emerge in systems of strongly correlated electrons \cite{wietek22}. Moreover, several moiré TMD systems are found to feature topologically non-trivial bands \cite{wu18b,tao23}, hence providing a platform to study the interplay of band topology and strong interactions \cite{cai23,zeng23}.

Recent experiments on WSe$_2$/WS$_2$ moiré heterostructures as well as twisted WSe$_2$/WSe$_2$ homobilayers have demonstrated Mott insulating states at half filling, as well as correlated insulating states at various fractional fillings \cite{regan20,li21}, corresponding to  generalized Wigner crystals, where longer-ranged Coulomb interactions lead to the localization of charges on self-organized lattices.

A highly topical open problem in this context is concerned with possible magnetic states that would arise at lowest temperatures by interactions lifting the residual spin degeneracies of charges localized on lattice sites \cite{pan20a,hu21,wietek22,motruk22}.
Since at some fractional fillings, the effective charge lattices correspond to frustrated lattices, these states may be prime candidates for the realization of highly sought-after quantum spin liquids \cite{savary2016quantum,broholm20}, featuring long-range entanglement and fractionalized excitations.

In this work, we focus on the moiré-Hubbard model with nearest-neighbor hopping $t$, onsite and longer-range Coulomb repulsions $U$ and $V_{ij}$ as an effective model for correlated electrons/holes on an emergent moiré triangular lattice, as applicable for $K$-valley moiré TMD heterostructures, such as WSe$_2$/WS$_2$ heterobilayers or twisted WSe$_2$ homobilayers \cite{wu18,pan20a,tang20}.
In principle, given a particular charge-ordered configuration (stabilized by onsite ($U$) and  longer-ranged ($V_{ij}$) repulsive interactions), one can perform a perturbative expansion in the hopping $t \ll U,V_{ij}$ to derive an effective strong-coupling Hamiltonian that lifts the spin degeneracy.
However, this procedure is cumbersome in practice, requiring separate perturbative expansions for each filling factor.
Further, at dilute fillings, spin-spin interactions may only be induced by processes at higher order in perturbation theory \cite{motruk22,seiba23}, and finding the ground states/phase diagrams of resulting spin Hamiltonians (often with multiple competing interactions) requires significant numerical efforts.
Such a program was undertaken recently by Motruk \textit{et al.} in Ref.~\onlinecite{motruk22}, where an effective spin model for pseudospin-1/2 degrees of freedom in the kagome charge crystal (at filling 3/4) was studied using density matrix renormalization group (DMRG) simulations, finding chiral spin liquid and kagome spin liquid phases.

In the paper at hand, we instead pursue a more integrated approach, aiming at a framework to simultaneously describe charge-ordered states at various fractional fillings and the concomitant magnetic states on top of these states.
To this end, we employ a slave-rotor representation, first introduced by Florens and Georges \cite{florens04}.
In this representation, each electron is fractionalized into a fermionic spinon (carrying spin, but no charge) and an $\mathrm{O}(2)$ rotor degree of freedom, with its angular momentum corresponding to the electronic charge.
Within the slave-rotor representation, the Hubbard model then becomes a model of interacting spinons and rotor degrees of freedom.
Making a mean-field approximation, this interacting problem can be split into a solvable free spinon Hamiltonian and a quantum XY model, with self-consistency equations coupling the two mean-field Hamiltonians.

Solving these self-consistency equations numerically allows us to map out phase diagrams, and characterize the nature of respective phases.
We summarize our main results below:
\begin{enumerate}
    \item We find various incompressible (Mott-insulating) states with charges forming emergent honeycomb ($\bar{n}=4/3,5/3$ filling) and kagome ($\bar{n}=5/4,7/4$) Wigner crystals on the (moiré) triangular lattice, as observed in previous experiments \cite{regan20}. These states are accessible by tuning chemical potential and/or the overall strength of repulsive electronic interactions (compared to kinetic energy scales).
    \item Distinct insulating states are separated by metallic (compressible) states, that are entered via first-order metal-insulator transitions. At fractional fillings, ``partially metallic'' states are found, where quasiparticles disperse on emergent lattices that are inherited from adjacent Wigner crystal phases, e.g., honeycomb or kagome sublattices of the triangular lattice.
    \item Considering incompressible (insulating) states, within our mean-field theory it is energetically preferable for the spinons to form dimerized states, corresponding to (possibly fluctuating) valence-bond solid (VBS) magnetic states that lift the remaining spin degeneracy of localized charges.
    We further find that some $\Uone$ spin liquid states are energetically competitive to these dimer states, such as 0-flux spinon Fermi surface $\Uone$ spin liquid and staggered $\pi$-flux $\Uone$ Dirac spin liquid on half-filling triangular charge crystal and 5/4 filling kagome charge crystal. We discuss the stability of such $\Uone$ spin liquid states on these different charge crystals.
 \end{enumerate}

The outline of this paper is as follows.
In Sec.~\ref{modelsymm} we briefly describe the generalized Hubbard model on the effective moiré triangular lattice, and detail the symmetry properties of the physical moiré system and the effective Hubbard model.
In Sec.~\ref{sec:slaverotmft}, we introduce the slave-rotor representation and mean-field approximation, and describe the solution of self-consistent equations for the decoupled free fermion and rotor Hamiltonian.
In Sec. \ref{Sec4}, we discuss results of our mean-field calculation and present phase diagrams as a function of chemical potential and interaction strength.
In Sec. \ref{Sec5}, we explore the physics beyond mean field theory, give arguments on the stability of different spin liquids and discuss possible experimental signatures.
A summary and outlook is given in Sec. \ref{Sec:summary}. 

\section{Moiré-Hubbard model} \label{modelsymm}

\subsection{Hamiltonian}

In this work, we are concerned with the moiré-Hubbard model on an effective triangular lattice (on moiré lattice scales), with the Hamiltonian
\begin{equation}
        H = H_t + H_U \label{eq:hubbard} \\
\end{equation}
where
\begin{subequations}\begin{align} 
    H_t &= 
    \varepsilon_0 \sum_{i,\sigma=\uparrow,\downarrow} c_{i,\sigma}^\dagger c^{\vphantom\dagger}_{i,\sigma}
    +
    \sum_{ij,\sigma=\uparrow,\downarrow} t_{ij,\sigma} c_{i,\sigma}^\dagger c^{\vphantom\dagger}_{j,\sigma}
    \label{eq:h-t} \\
    H_U &= \frac{U}{2} \sum_i (n_i-1)^2 + \frac{1}{2}\sum_{ij}V_{ij} (n_i - 1) (n_j-1). \label{eq:h-u}
\end{align}\end{subequations}
Here, $\varepsilon_0$ is the onsite energy, $i,j$ denote lattice sites of the effective moiré triangular lattice, $t_{ij,\sigma}$ corresponds to a (possibly complex) spin-dependent hopping amplitude, $U$ is the onsite Coulomb repulsion, and $V_{ij}$ is the long range Coulomb interaction.
As discussed further below, we will mostly focus on truncating long-ranged Coulomb interaction to nearest neighbor $V$ and next-nearest neighbor $V^{\prime}$ for simplicity, but write $V_{ij}$ for generality.
The total electron number $n_i = n_{i,\uparrow} + n_{i,\downarrow}$.

Note that we have defined the interaction term \eqref{eq:h-u} so that $\varepsilon_0=0$ corresponds to half-filling.
$H_U$ is related (up to a constant) to the conventional form $U\sum_i n_{i,\uparrow} n_{i,\downarrow} + V \sum_{\langle ij \rangle} n_i n_j + V^{\prime} \sum_{\langle\langle ij \rangle\rangle}n_in_j$ by a redefinition of the onsite energy $\epsilon_0 \to \epsilon_0 - U/2 - 6(V+V^{\prime})$.

As written, the Hamiltonian is agnostic regarding specific material realizations.
A general principle that gives rise to such effective moiré-Hubbard Hamiltonians consists in determining the band structure that arise when holes near the valence-band maxima (VBM) experience a slowly varying periodic moiré potential (in heterobilayers, induced by a second layer with incommensurate lattice geometry), or a periodically varying interlayer hybridization (in twisted homobilayer systems).
Considering nearly-flat and well-isolated moiré bands, one may then construct appropriate localized Wannier orbitals, with their overlaps giving rise to the tight-binding dispersion $H_t$, and $H_U$ is obtained from projecting Coulomb interactions onto these localized orbitals.
The locations of the centers of these Wannier orbitals therefore determine the effective lattice geometry in Eq.~\eqref{eq:hubbard}.
In TMDs with $\bvec{K}$-valley VBM, strong spin-orbit coupling leads to a locking of spin and valley degrees of freedom near the Fermi level, so that quasiparticles in the effective moiré-Hubbard model carry a single (combined) $S_\mathrm{eff}=1/2$ spin-valley degree of freedom (pseudospin). 

We briefly discuss possible material realizations:
\begin{enumerate}
    \item Heterobilayers such as WSe$_2$/MoSe$_2$ \cite{wu19} or WSe$_2$/WS$_2$\cite{tang20}: the topmost moir{\'e} band originates from the $\bvec{K}$/$\bvec{K}'$ valley valence electrons of the WSe$_2$ layer, experiencing a triangular moir{\'e} potential modulated by the WS$_2$ or MoSe$_2$ layer. Wannier centers constructed from this moiré band are found to form an effective triangular lattice \cite{wu18}.
    \item Twisted homobilayers such as twisted bilayer WSe$_2$ \cite{pan20b,dean20}: the $\bvec{K}$/$\bvec{K}'$ valley valence bands from both layers hybridize to generate the (topologically trivial) $\bvec{K}$/$\bvec{K}'$ valley moir{\'e} bands, respectively. Wannier centers are found to form a triangular lattice, coinciding with sites in the moiré structure where the metal atoms in the two layers are aligned \cite{pan20a}.
\end{enumerate}
For both realizations, the topmost moir{\'e} bands are doubly degenerate and related to each other by time reversal symmetry, corresponding to the pseudospin-1/2 degeneracy. 

For some twisted TMDs, most prominently twisted bilayer MoTe$_2$ \cite{wu18b}, Wannier states for the topmost moiré bands are found to form an effective honeycomb superlattice, with pseudospin-1/2--dependent intralayer hopping giving rise to an effective realization of the Kane-Mele model. The interplay of band topology and strong interactions has recently received immense attention, following experimental reports of fractional quantum anomalous Hall states \cite{zeng23,cai23}.

In the following, we will focus on moiré TMD systems well-described by effective triangular Hubbard models, for which generalized Wigner crystal states have been observed experimentally \cite{li21}.
We stress that that in principle, our slave-rotor mean-field study as presented in Sec.~\ref{sec:slaverotmft} can be straightforwardly applied to appropriate Kane-Mele-Hubbard models, which is an interesting avenue left for further study.
However, we note that pseudospin-dependent complex next-nearest neighbor hoppings give rise to Dyzaloshinskii-Moriya interactions in the strong-coupling limit which can be expected to stabilize (non-collinear) magnetic order rather than spin-liquid phases \cite{wietek22}.

\subsection{Symmetries} \label{sec:symm}

Moiré heterostructures have distinct microscopic symmetries.
TMD monolayers possess $C_{3v}$ symmetry, with a vertical reflection plane parallel to the links of the effective honeycomb lattice.
For \emph{(twisted) heterobilayers}, this reflection symmetry is broken, and the $C_{3v}$ symmetry is reduced to a $C_3$ symmetry.
\emph{Twisted homobilayers} have $D_3$ symmetry which is generated by $C_3$ rotations as well as $C_{2}$ rotations around an in-plane axis which swaps the top and bottom layers. 
Note that vertical displacement field, e.g., introduced by gate voltages, would break the layer pseudo-spin symmetry and reduce it to a $C_3$ symmetry.
For both systems, time reversal symmetry is preserved, connecting the $\bvec{K}$(spin up) and $\bvec{K}'$(spin down) degrees of freedom.

The effective Hubbard model Eq.~\eqref{eq:hubbard} is constructed by projecting the repulsive Coulomb interactions to the lowest energy (flat) bands derived from continuum models for $\bvec{K}$-valley moiré TMD \cite{pan20a,pan20b}.
Crucially, in Refs.~\cite{pan20a,pan20b} moiré potentials were truncated beyond the lowest harmonics (i.e.~restricting to Fourier components corresponding to the first six moiré reciprocal lattice vectors).
As we detail in App.~\ref{sec:acc-symm}, this truncation leads to the emergence of an accidental inversion symmetry of the moiré-Bloch wavefunctions, and thus also of the effective Hubbard model for the respective Wannier states.

Now we comment on the validity of the lowest harmonics approximation, following the original argument in the Bistritzer-MacDonald paper \cite{MacDonald11} for twisted bilayer graphene. We expect the interlayer tunneling amplitude $t_{\bm{q}}$ at momentum $\bm{q}$ to drop rapidly on the reciprocal lattice vector scale. For example, based on Ref. \cite{devakul2021magic}, WSe$_2$ has interlayer separation $6.7$\AA $\leq d_{\perp}\leq 7.1$\AA, which exceeds the intralayer lattice constant $a=3.28$\AA~by more than a factor of two. Because the real-space hopping $t(\bm{r})$ varies with three-dimensional separation $\sqrt{d_{\perp}^2+r^2}$, $t_{\bm{q}}$ decreases rapidly for $qd_{\perp}>1$. 

\section{Slave-rotor mean-field theory} \label{sec:slaverotmft}

We seek an integrated description of metal-insulator transitions (at zero temperature) and the concomitant formation of charge crystals at certain fractional fillings, and the magnetic states in the incompressible regimes (with localized charges).
To this end, we employ a slave-rotor representation to explicitly separate the electrons' spin and charge degrees of freedom.
While in an exact rewriting these are strongly coupled, we can make a mean-field approximation to obtain separate spin and charge Hamiltonians, coupled via self-consistency equations.

\subsection{Slave-rotor representation}

Following Ref.~\onlinecite{florens04}, we split electrons at each site into fermionic chargeless spinons and a single on-site O(2) rotor degree of freedom. The rotor is used to represent the phase degree of freedom $\theta_i$, conjugate to the total charge at site $i$, identified as the rotor's angular momentum $\hat{L}_i =-\iu \partial/\partial \theta_i$.
The electron creation operator at site $i$ is rewritten as
\begin{equation}
    c_{i,\sigma}^{\dagger}=f_{i,\sigma}^{\dagger} \eu^{\iu \theta_i} \label{eq:creat_operat}
\end{equation}
where the spinon $f_{i,\sigma}^{\dagger}$ has the same spin/orbit flavor as the electron, and $\eu^{\iu \theta_i}$ raises the angular momentum of the rotor by one unit.
In other words, creating an electron amounts to creating a spinon and raising the angular momentum (total charge) by one at the same time.
This rewriting enlarges the local Hilbert space and thus introduces redundant degrees of freedom.
Therefore, a constraint is imposed that the number of spinons match the total charge,
\begin{equation}
    \hat{L}_i =\sum_{\sigma}(f_{i,\sigma}^{\dagger}f^{\vphantom\dagger}_{i,\sigma}-1/2) \label{eq:constr}.
\end{equation}
Here, we choose the convention that the rotor quantum number $L_i=0$ corresponds to half filling, e.g., for electrons with spin-1/2, ${L_i}=0$ implies that there is exactly one electron at site $i$.

We note that the representation Eq.~\eqref{eq:creat_operat} is an exact rewriting if the constraint Eq.~\eqref{eq:constr} is enforced on each site, for example by means of a Gutzwiller projection.
However, since there is only limited analytical understanding of projected wavefunctions, and their evaluation requires significant numerical efforts, we instead henceforth will enforce the constraint Eq.~\eqref{eq:constr} \emph{on average}. 

The merit of the slave-rotor representation lies in the fact that Coulomb repulsion is only dependent on the charge quantum number, and we can thus replace the four-fermion interaction terms $H_U$ [see Eq.~\eqref{eq:h-u}] by terms quadratic in the rotor's angular momentum.
Specifically, considering an atomic Hamiltonian with some onsite energy level $\varepsilon_0$ and on-site Coulomb repulsion $U$, we can write
\begin{align}
    H_\mathrm{at} &= \sum_\sigma \varepsilon_0 c^\dagger_{\sigma} c^{\vphantom\dagger}_\sigma + \frac{U}{2} ( n-1)^2  \nonumber\\
    &= \sum_\sigma \varepsilon_0 f^\dagger_\sigma f^{\vphantom\dagger}_\sigma + \frac{U}{2} \hat{L}^2
\end{align}
where we drop an overall numerical constant.
We now generalize to the Hubbard model in Eq.~\eqref{eq:hubbard}.
Again using the slave-rotor representation in Eq.~\eqref{eq:creat_operat}, and replacing $\hat{L}_i = n_i - 1$, the Hubbard Hamiltonian can be expressed in terms of spinons and rotors as
\begin{align} \label{eq:h-slaverotorrep}
    H &=
    -\sum_{i,\sigma} \mu f^{\dagger}_{i,\sigma}f^{\vphantom\dagger}_{i,\sigma}
    +\sum_i\frac{U}{2}\hat{L}_i^2
    +\frac{1}{2} \sum_{ij}V_{ij} \hat{L}_i\hat{L}_j    \notag \\
    &-\sum_{ij,\sigma}t_{ij,\sigma}
    f^{\dagger}_{i,\sigma}f^{\vphantom\dagger}_{j,\sigma}
    \eu^{\iu (\theta_i-\theta_j)}.  
\end{align}
Here, we have replaced the onsite energy $\epsilon_0$ by a chemical potential,
\begin{equation}
    \epsilon_0 = - \mu,
\end{equation}
which is an experimentally accessible tuning parameter (via electrostatic gating) \cite{tang20}.
In the following, we will therefore work in the grandcanonical ensemble rather than at fixed particle number.

The kinetic term of the Hubbard model has become a coupling between spinon and rotor degrees of freedom in Eq.~\eqref{eq:h-slaverotorrep}.
In principle, the constraint Eq.~\eqref{eq:constr} should be imposed on each site.

\subsection{Mean-field decoupling of spinons and charge rotors} \label{subsec: mf decouple}
The fermionic spinons and rotor degrees of freedom interact via the ``correlated hopping'' in Eq.~\eqref{eq:h-slaverotorrep}, preventing an exact solution of the model. 
To make progress, here we perform a mean-field decoupling of the interaction term,
\begin{multline}
    f^{\dagger}_{i,\sigma} f^{\vphantom\dagger}_{j,\sigma}\eu^{\iu (\theta_i-\theta_j)}\rightarrow \left<f^{\dagger}_{i,\sigma}f^{\vphantom\dagger}_{j,\sigma}\right>\eu^{\iu (\theta_i-\theta_j)}\\
+f^{\dagger}_{i,\sigma} f^{\vphantom\dagger}_{j,\sigma}\left<\eu^{\iu (\theta_i-\theta_j)}\right>
-\left<f^{\dagger}_{i,\sigma}f^{\vphantom\dagger}_{j,\sigma}\right> \left<\eu^{\iu (\theta_i-\theta_j)} \right>,
\end{multline}
where $\langle \dots \rangle$ denotes an expectation value with respect to the ground state of the respective mean-field Hamiltonian.

We also add a Lagrange multiplier field $h_i$ to impose the constraint Eq.~\eqref{eq:constr}.
The Hamiltonian Eq.~\eqref{eq:h-slaverotorrep} then splits into separate Hamiltonians for the fermionic spinons and O(2) quantum rotors,
\begin{align}
    H_f&=\sum_{i,\sigma}(-\mu-h_i)f^{\dagger}_{i,\sigma}f^{\vphantom\dagger}_{i,\sigma}
    -\sum_{ij,\sigma}t_{ij,\sigma}^{\text{eff}}f^{\dagger}_{i,\sigma}f^{\vphantom\dagger}_{j,\sigma}   \\
    H_{\theta}&=\sum_{i} \frac{U}{2}\hat{L}_i^2+h_i\hat{L}_i
    + \sum_{ij}\frac{1}{2} V_{ij}\hat{L}_i\hat{L}_j
    -K_{ij}\eu^{\iu (\theta_i-\theta_j)} ,  \label{eq:Hr}
\end{align}
where the effective hopping $t_{ij,\sigma}^\mathrm{eff}$ for the (free) fermionic spinons and the coupling of quantum rotors $K_{ij}$ are related to the mean-field parameters, and are to be determined self-consistently.
Explicitly, the coupled self-consistency relations read
\begin{align} 
t_{ij,\sigma}^{\text{eff}}&=t_{ij,\sigma}\left\langle\eu^{\iu (\theta_i-\theta_j)}\right\rangle    \label{eq:teff} \\
K_{ij}&=\sum_{\sigma}t_{ij,\sigma}\left\langle f^{\dagger}_{i\sigma}f^{\vphantom\dagger}_{j\sigma}\right\rangle. \label{eq:Keff}
\end{align}
Further, the parameters $h_i$ must be (implicitly) determined to satisfy the average constraint for matching the filling of spinons to each site's charge,
\begin{equation}
\left<\hat{L}_i\right>=\sum_{\sigma}\left[\left<f^{\dagger}_{i,\sigma}f^{\vphantom\dagger}_{i,\sigma}\right>-1/2\right]. \label{eq:const_ave}
\end{equation}

\subsection{Solution of rotor Hamiltonians}

While the free fermion Hamiltonian is easily diagonalized by means of a unitary transformation in momentum space, the Hamiltonian $H_\theta$ of interacting O(2) rotors (i.e.~a \emph{quantum} XY model) evades such an exact solution.

Instead, we will make physically motivated approximations to the rotor correlation expectation value $\left<\eu^{\iu (\theta_i-\theta_j)}\right>$ which characterizes distinct phases of the quantum XY model, and in the present context then determine the effective spinon dispersion.
We discuss two distinct regimes below.

\subsubsection{Metallic (compressible) states} \label{sec:metal}

The rotor acquiring a non-zero expectation value $\langle \eu^{\iu \theta_i}\rangle \equiv \sqrt{Z_i} \neq 0$ can be understood to be analogous to the condensation of a bosonic ladder operator $\braket{b^\dagger_i}$, giving rise to a superfluid phase for the bosonic degrees of freedom.
This phase is characterized by off-diagonal long range order of the rotor correlator at long distances, i.e.~$\lim_{|i-j| \to \infty} \left< \eu^{\iu (\theta_i - \theta_j)} \right> = \left< \eu^{\iu \theta_i} \right> \left< \eu^{-\iu \theta_j} \right>$.
We can access this phase on a mean-field level by factorizing the correlator $\left< \eu^{\iu (\theta_i - \theta_j)} \right> \approx \left< \eu^{\iu \theta_i} \right> \left< \eu^{- \iu \theta_j} \right>$.
This implies that the effective spinon hopping [cf. Eq.~\eqref{eq:teff}] can be written as
\begin{align}
    t_{ij,\sigma}^{\mathrm{eff}}&=t_{ij\sigma}\braket{\eu^{\iu \theta_i}}\braket{\eu^{-i\theta_j}}    \notag \\
    &=t_{ij\sigma}\sqrt{Z_i}\sqrt{Z_j}    \label{eq:teff_m}
\end{align}
where we assume $\braket{\eu^{\iu \theta_i}}$ is real, which should be expected if the time reversal symmetry is unbroken, and then $\braket{\eu^{\iu \theta_i}}=\braket{\eu^{-i\theta_i}}=\sqrt{Z_i}$ from hermiticity. The notation $\sqrt{Z_i}$ is used so that $Z_i$ would have the meaning of spectral weight, see below. The nonzero expectation value of the phase operator indicates that the rotor's angular momentum, and thereby the electronic charge, is no longer a good quantum number and thus the system is in a metallic (compressible) state.
Especially at commensurate fillings, upon increasing Coulomb repulsion, $Z_i$ decreases continuously to zero, which is the well-known Mott transition as demonstrated in Ref. \cite{florens04}.

Within the slave-rotor formalism, the electronic Green's function $G^{(c)}$ is given by
\begin{equation}
    G^{(c)}_{ij}(\tau-\tau') = G^{(f)}_{ij}(\tau-\tau') \langle \eu^{-\iu [\theta_i(\tau) - \theta_j(\tau')] } \rangle.
\end{equation}
In the metallic states, the rotor degrees of freedom are long-range ordered, and in the mean-field approximation we can read off the spectral weights of the electronic quasiparticles as $Z_{ij} = \sqrt{Z_i} \sqrt{Z_j}$
where $\sqrt{Z_i} = \langle \eu^{\iu \theta_i} \rangle$. Note that here, the spinon bands contribute unity spectral weight, such that the wavefunction renormalization of the electronic quasiparticles is determined by the rotor degrees of freedom.

To explicitly solve the self-consistency equations, we note that with Eq.~\eqref{eq:teff_m}, the rotor Hamiltonian Eq.~\eqref{eq:Hr} can be written as
\begin{align}
    H_{\theta}&\approx \frac{U}{2} \sum_i  \hat{L_i}^2 + \frac{1}{2}\sum_{ij}V_{ij}\hat{L_i}\hat{L_j}+\sum_i h_i\hat{L_i}
    \notag \\
    &-\sum_i \Big( \sum_{j} K_{ij}\sqrt{Z_j} \Big)\big(\eu^{\iu\theta_i}+\eu^{-\iu\theta_i}\big)+\text{const.}  \label{eq:Hr_metal}
\end{align}
In line with our site-factorized treatment of the rotor kinetic energy, we also decouple the long range Coulomb interaction as
\begin{equation}
    \sum_{ij}V_{ij}\hat{L_i}\hat{L_j}\approx
    \sum_{ij}V_{ij}\left(\hat{L_i}\braket{\hat{L_j}}+\braket{\hat{L_i}}\hat{L_j}
    -\braket{\hat{L_i}}\braket{\hat{L_j}}\right)
\end{equation}
Then, the rotor Hamiltonian can be reduced to a sum of decoupled single-site rotor (mean-field) Hamiltonians
\begin{align}
    H_{\theta}^\mathrm{MF}&= \sum_i\Bigg[\frac{U}{2}\hat{L}_i^2+\Big(\sum_{j} V_{ij} \braket{\hat{L_j}}
    +h_i \Big)\hat{L_i}\Bigg] \notag \\
    &-\sum_i\Big(\sum_{j} K_{ij}\sqrt{Z_j} \Big)\big(\eu^{\iu\theta_i}+\eu^{-\iu\theta_i}\big)+\mathrm{const.}
\end{align}
Given a set of $K_{ij}$, the mean-field Hamiltonian $H_\theta^\mathrm{MF}$ can now be readily solved, where $\braket{\hat{L}_i}$ and $\sqrt{Z_i}$ are to be determined self-consistently -- the corresponding ground-state expectation values $\braket{\eu^{\iu \theta_i}}$ then determine $t^\mathrm{eff}_{ij,\sigma}$, which serves as an input for the solution of the fermionic spinon Hamiltonian, to obtain the value of $K_{ij}$ for the next iteration.

\subsubsection{Insulating (incompressible) states} \label{sec:insulating states}

We characterize insulating states by vanishing of the respective quasiparticle weights which attains when the phase operator expectation values $\braket{\eu^{\iu \theta_i}} = 0$.
In this case, $L$ would be quantized to be integers, giving rise to zero compressibility $\partial n /\partial \mu=0$.
When $\braket{\eu^{\iu \theta_i}}=\sqrt{Z_i}=0$, there is no long-range order for the rotor degrees of freedom, but we stress that this does not necessarily lead to $\braket{\eu^{\iu (\theta_i-\theta_j)}}=0$: A simple site-factorized mean-field treatment of the quantum XY model (as suggested for metallic states above) is incapable of correctly producing such finite (short-range) rotor correlations.
Instead, we obtain the expectation value of this operator from Eq.~\eqref{eq:Hr} by the Hellmann-Feynman theorem
\begin{equation}
    \braket{\eu^{\iu (\theta_i-\theta_j)}}
    =-\frac{\partial \braket{H_{\theta}}}{\partial K_{ij}},
    \label{eq:HF_thm}
\end{equation}
which makes explicit that in general rotor correlations do not vanish, since the energy expectation value will depend on $K_{ij}$.
This implies that in insulating states with vanishing quasiparticle weights $Z=0$, the spinon in general still has nonzero hopping and disperses.
Such spin-charge separation is typical of spin liquids.

In this work, we employ canonical perturbation theory to calculate the ground state energy $\langle H_\theta \rangle$ of the rotor Hamiltonian Eq.~\eqref{eq:Hr} and the rotor correlation from Eq.~\eqref{eq:HF_thm} when the onsite mean-fields $\braket{\eu^{\iu \theta_i}}=0$ vanish.
This perturbative expansion is controlled if the rotor-rotor coupling $K_{ij}$ is small compared to the repulsive $U,V$ interactions.
To this end, we take the (solvable) Hamiltonian for the angular momenta as the unperturbed Hamiltonian
\begin{equation}
    H_{\theta}^{(0)}=\sum_i\frac{U}{2}\hat{L}_i^2
    +\frac{1}{2} \sum_{ij}V_{ij}\hat{L}_i\hat{L}_j+\sum_i h_i\hat{L}_i,
\end{equation}
with the perturbation
\begin{equation}
   H_{\theta}^{\prime}=\sum_i\delta h_i \hat{L}_i
    -\sum_{ij}K_{ij}\eu^{\iu (\theta_i-\theta_j)} 
\end{equation}
where we included $\delta h_i$ as a possible change of Lagrange multipliers to ensure the constraint Eq.~\eqref{eq:const_ave} remains satisfied.
Then, the ground-state energy up to second order in $K/U$ reads
\begin{equation}
    E_0=E_0^{(0)}+\sum_i\delta h_i L_i
    +\sum_{ij}\frac{K_{ij}K_{ji}}{E_0^{(0)}-E_{ij}^{(0)}}
    +\mathcal{O}((K/U)^3).     \label{eq:E_pert}
\end{equation}
Here, $E_{ij}^{(0)}$ corresponds to the energy of the configuration that one unit charge is moved from site $j$ to site $i$, with respect to the unperturbed ground configuration. 
From Eq.~\eqref{eq:HF_thm} we can see that $\braket{\eu^{\iu (\theta_i-\theta_j)}}$ and thus $t_{ij,\sigma}^{\text{eff}}$ would be proportional to $K_{ij}$ to the lowest order.

Eq.~\eqref{eq:HF_thm} and \eqref{eq:E_pert} correspond to a set of self-consistent equations, now accounting for perturbative corrections. From these we can solve for $K_{ij}$ and other parameters.
There is always a trivial solution $K_{ij}=0$, which is the normal insulating state. When $K_{ij}\neq 0$, we get nonzero spinon hoppings, while the system is still incompressible because the charge is quantized and conserved ($\braket{\eu^{\iu \theta}}=0$).

\subsubsection{Classification of spin liquid states}    \label{sec: classif}

When the $K_{ij}$ are not zero in insulating states, we can get a larger set of solutions than the metallic case, which fall into different universality classes of spin liquids.
In fact, in the decomposition Eq.~\eqref{eq:creat_operat}, we have a $U(1)$ gauge redundancy
\begin{align}
    f^{\dagger}_{i,\sigma}&\rightarrow 
    \eu^{-i\varphi_i}f^{\dagger}_{i,\sigma}  \notag \\
    \theta_i &\rightarrow \theta_i + \varphi_i \label{eq:U1_tran}
\end{align}
After this transformation, the electron operators and thus the \emph{physical} Hamiltonian remain invariant. However, given mean-field Hamiltonians $H_f$ and $H_\theta$ will in general not be invariant under these transformations.
Equivalently, one observes that even though a physical wavefunction for a spin liquid may preserve all physical symmetries (space group symmetries, spin rotation symmetry and time reversal symmetry), a corresponding mean-field Hamiltonian is only invariant if those symmetry operations are supplemented by appropriate gauge transformations as in Eq.~\eqref{eq:U1_tran}.
In other words, symmetries of mean-field ansaetze are realized projectively.
Mean-field ansaetze corresponding to distinct physical states can be classified by their respective projective symmetry groups (PSG), as introduced in Ref.~\cite{wen02}.

To be more explicit, consider a space group transformation $U$ under which the physical state $\left|\Psi_{\text{Phys}}\right>$ is invariant.
Before projection, the mean field state $\left|\Psi^{(K_{ij})}\right>$ may not transform trivially under $U$, since $K_{ij}$ and $K^{\prime}_{ij}$ related by a $\Uone$ gauge transformation [Eq.~\eqref{eq:U1_tran}] correspond to actually the same physical state.
Invariance of the mean field ansatz $K_{ij}$ is achieved by combining $U$ with a $\Uone$ gauge transformation
\begin{equation}
    G_U U(K_{ij})=K_{ij} \label{eq:PSG}
\end{equation}
where the physical operation $U$ maps the spatial index $i$ to some other index $U(i)$, and $G_U$ is an appropriately chosen $\Uone$ gauge transformation,
\begin{align}
    G_U: &f^{\dagger}_{i\sigma}\rightarrow \eu^{-\iu \varphi_{U(i)}} f^{\dagger}_{i\sigma} \notag \\
    &\eu^{\iu \theta_i}\rightarrow \eu^{\iu \varphi_{U(i)}}\eu^{\iu \theta_i} \notag \\
    &K_{ij}\rightarrow \eu^{-\iu(\varphi_{U(i)}-\varphi_{U(j)})}K_{ij} \notag \\
    &t_{ij,\sigma}^{\text{eff}}\rightarrow
    \eu^{\iu(\varphi_{U(i)}-\varphi_{U(j)})} t_{ij,\sigma}^{\text{eff}}
\end{align}
The set of $G_UU$ that leaves $K_{ij}$ invariant is referred to as the invariant
PSG.
Different PSG entail distinct $K_{ij}$ ansaetze, and thus we can classify the mean field ansaetze and corresponding \emph{physical} states by their PSG realizations.

The subgroup of invariant PSG that is a pure gauge group is called invariant gauge group (IGG), and the physical symmetry group is hence given by $\mathrm{SG} = \mathrm{PSG}/\mathrm{IGG}$.
Here in the insulating case, a global $\Uone$ transformation leaves $K_{ij}$ and $t_{ij,\sigma}^{\text{eff}}$ invariant
\begin{align}
    f^{\dagger}_{i,\sigma}&\rightarrow 
    \eu^{-i\varphi}f^{\dagger}_{i,\sigma}  \notag \\
    \eu^{\iu \theta_i}&\rightarrow \eu^{\iu \varphi}\eu^{\iu \theta_i}
\end{align}
where $\varphi$ is site independent. Therefore, the slave-rotor representation has a $\Uone$ IGG.

Then, we can ask how many gauge-inequivalent classes of spin liquids are allowed if we demand the full physical symmetry. The structure of the symmetry group imposes an algebraic constraints on the PSG. For example, demanding that translations along the two principal axis of a lattice commute implies
\begin{equation}
    T_1^{-1}T_2T_1T_2^{-1}=\mathcal{I} \label{eq:phys translation}
\end{equation}
Then, the PSG Eq.~\eqref{eq:PSG} with a $\Uone$ IGG should satisfy
\begin{equation}(G_{T_1}T_1)^{-1}G_{T_2}T_2G_{T_1}T_1(G_{T_2}T_2)^{-1}
    =G\in U(1)
\end{equation}
The identity of right hand side of Eq.~\eqref{eq:phys translation} is now relaxed to an element of $\Uone$ IGG since it keeps the mean-field ansaetze invariant. Listing out all the relations of the symmetry group, we can get a finite number of distinct PSGs allowed by these algebraic constraints, called algebraic PSG.
Since ansaetze $K_{ij}$ are distinguished by their PSG realizations, we arrive at a finite number of choices, which are not related to each other by a pure gauge transformation, and thus fall into different classes.
We can therefore focus on particular mean-field ansaetze $K_{ij}$ that correspond to distinct PSG, under a given symmetry requirement.

Above arguments and PSG classifications in general employ the full symmetry group of the lattice on which spin degrees of freedom reside.
However, in this work, we are in particular focused on charge-ordered states that may form in the generalized Hubbard model.
At fractional fillings, charges then reside on effective sublattices of the triangular lattice (e.g.~honeycomb lattice at 4/3 filling or kagome lattice at 5/4 filling).
For these states, a similar PSG analysis can be applied based on the symmetry groups of the respective charge-ordered states (that spontaneously break translational/rotational symmetries of the parent triangular lattice).
Moreover, if we allow a breaking of rotational symmetry, nematic states are possible, where the amplitudes of $K_{ij}$ differ on bonds of different orientation.
These also include dimer states that are formed if all the nonzero $K_{ij}$ bonds are disconnected, corresponding to VBS states.

In the following, based on the discussion above, we will consider distinct invariant PSG ansaetze that have previously been found to be energetically competitive on various charge crystals, solve the self-consistent equations respectively, and compare their respective energies.

We mention that also in the metallic phase, the slave-rotor Hamiltonian before mean-field decoupling formally enjoys a $\Uone$ gauge redundancy.
But the phase operator $\eu^{\iu \theta} \rightarrow \langle \eu^{\iu \theta} \rangle \neq 0$ acquiring a finite expectation value implies that the IGG here is just a trivial group with identity as the only element, because $\braket{\eu^{\iu \theta}}$ would change under any transformation $\theta_i \to \theta_i + \phi_i$. This trivial IGG will result in only one solution allowed where $K_{ij}$ are the same on all bonds (a uniform solution), if the full physical symmetry is preserved: This metallic state does not possess an intrinsic gauge structure and corresponds to a confining phase.

We further note that dimer states which have $\langle f^\dagger_{i,\sigma} f^{\vphantom\dagger}_{j,\sigma} \rangle \neq 0$ on disjoint bonds $\langle ij \rangle$ possess \emph{on the mean-field level} a $\Uone$ IGG. However, the disconnected nature of the spinon hopping, all fermionic degrees of freedom are gapped (i.e.~spinons are localized on bonds, forming spin singlets after projection).
Pure (compact) $\Uone$ gauge theory in 2+1-dim. is unstable \cite{poly75,polyako77}, and thus the dimer phase must be a confining phase of matter.

\section{Mean-field phases} \label{Sec4}

\subsection{Overview}
To map out phase diagrams, we numerically solve the self-consistency equations Eqs.~\eqref{eq:teff} and \eqref{eq:Keff} in Sec.~\ref{subsec: mf decouple}, as a function of chemical potential $\mu$ and interaction strength $U$.
We find that phase diagrams depend significantly on the presence and nature of longer-ranged repulsive interactions $V_{ij}$ in Eq.~\eqref{eq:h-u} due to Coulomb interactions between charges.

Typically, the Coulomb interaction in bulk solids is efficiently screened (leading to an exponential decay with distance), justifying the approximation of repulsive interactions as an onsite (contact) interaction.
However, in two-dimensional moir{\'e} heterostructures, screening is significantly weaker, and the moiré-induced quenching of kinetic energy scales implies that extended repulsive interactions are no longer negligibly small \cite{mak22}. As we shall see below (and discuss in Appendix~\ref{app:long-range-coulomb}), extended repulsive interactions (beyond nearest-neighbor) are necessary for reproducing some of the more complex generalized Wigner crystals at certain filling factors.

In fact, the screening length in moiré TMD heterostructures can be controlled via the choice of and distance to metallic screening layers (which also act as gate electrodes).
Modelling the screening via the method of image charges, the electron-electron interaction potential is $U(r)=(e^2/\epsilon)[r^{-1}-(r^2+D^2)^{-1/2}]$, where $D$ is the vertical distance between the metallic layer and TMD bilayer \cite{wu18}.

With this in mind, in this study, we for simplicity consider two distinct cases of next-nearest neighbor repulsion $V^{\prime}$:
\begin{enumerate}
    \item $V^{\prime}=0$, corresponding to short-ranged (truncated beyond nearest-neighbor repulsion $V$) interactions due to strong screening.
    \item $V^{\prime}=\frac{1}{\sqrt{3}}V$, roughly motivated by a $1/r$ decaying Coulomb repulsion, such that $V^{\prime}/V$ is inversely proportional to distance ratio.
\end{enumerate}
In both cases, we neglect repulsions beyond next-nearest neighbors for simplicity, and the ratio of nearest neighbor to on-site Coulomb repulsion $V/U$ is fixed to be 1/4 \cite{pan20a,t_li21,xu20}.
We expect that in realistic systems, $V^{\prime}/V$ should take a value between the two cases discussed above, depending on microscopic details.
Note that here we also drop hopping amplitudes beyond nearest neighbors since the Wannier states on the moiré lattice scale are exponentially localized, and accordingly longer-ranged hopping has been found to be negligible compared to the strong Coulomb interactions in moir{\'e} TMDs \cite{t_li21,xu20}.

\begin{figure}
  \centering
  \subfloat[$\sqrt{3}\times\sqrt{3}$ unit cell]{\includegraphics[width=0.4\linewidth]{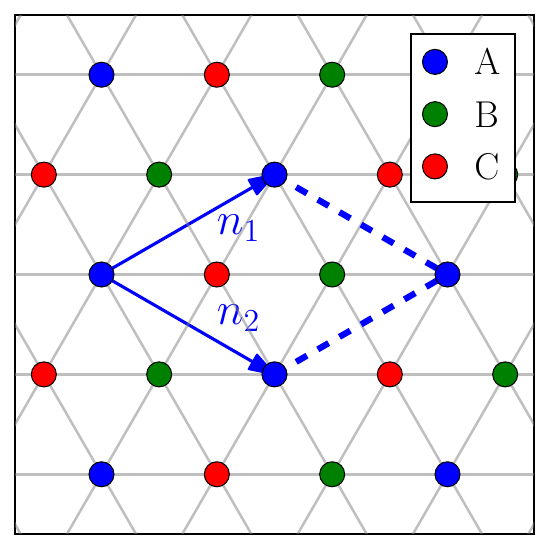}\label{fig:3-site uc}}
  \subfloat[$2\times2$ unit cell]{\includegraphics[width=0.4\linewidth]{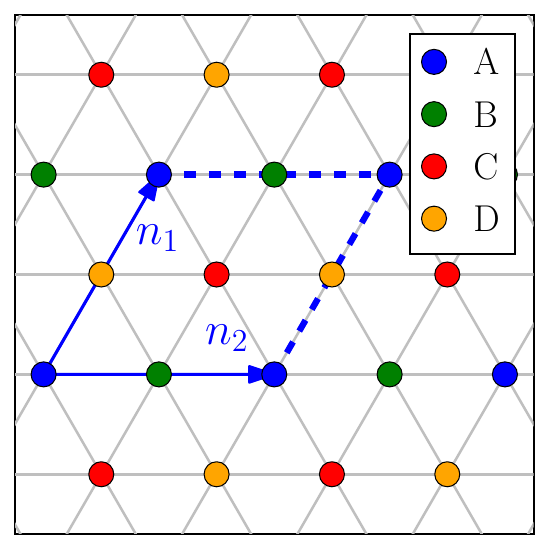}\label{fig:4-site uc}}
  \caption{Unit cells for 3-site ansatz and 4-site ansatz.}
  \label{fig:unit cell}
\end{figure}
\begin{figure*}
    \centering
    \includegraphics[width=\textwidth]{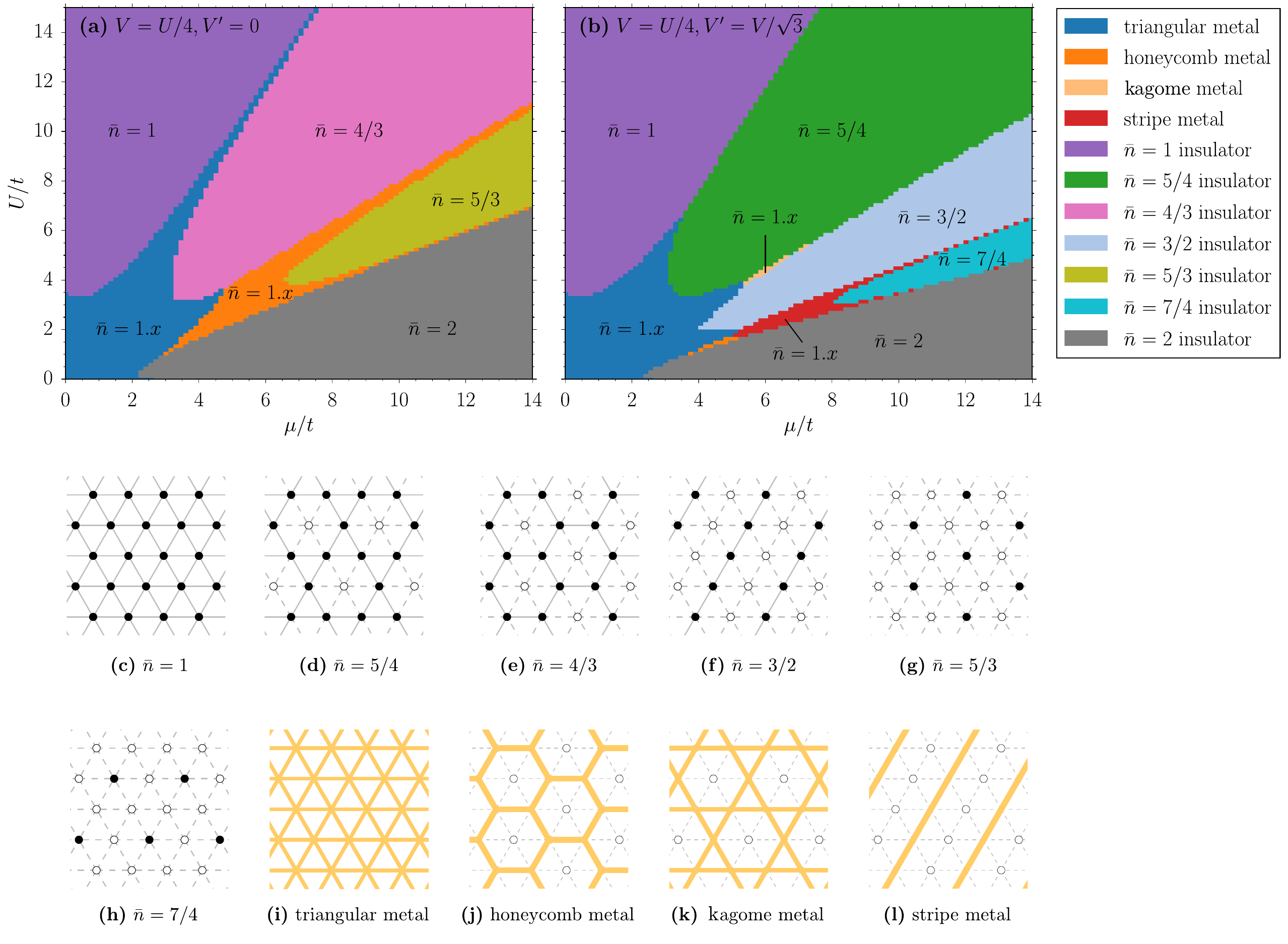}
    \caption{Mean field phase diagrams in the $\mu-U$ plane, for nearest-neighbor repulsion $V=U/4$, and two representative values of next-nearest neighbor repulsion $V'$, (a)  $V'=0$ and (b) $V'=V/\sqrt{3}$. Average particle number $\bar{n}$ is a commensurate fractional number for insulating phases, and $\bar{n}=1.x$ denotes incommensurate filling for metallic states conducting on different sublattices. 
    (c),~(d),~(e)~and~(f)~schematic depiction of various charge orders (corresponding to generalized Wigner crystals) found for correlated insulating states at fractional filling. Filled and empty circles correspond to  half-filled and doubly-occupied sites. On solid lines spinons may have non-zero hopping, whereas bonds denoted by dashed lines do not contribute to the connectivity of the lattice. 
    (i),~(j),~(k)~and~(l)~schematic depiction of metallic states, with charges dispersing on different sublattices of the parent moiré triangular lattice. Orange shading indicates bonds on which charges disperse, and empty circles correspond to doubly occupied sites.%
    }
    \label{fig:both-pd}
\end{figure*}

Considering only nearest-neighbor repulsion, a classical analysis shows that all possible charge crystals on the triangular have at most three distinct sublattices, so three inequivalent sublattice sites with possibly different $L_i, Z_i$ and $h_i$ are needed, assuming a translational symmetry with respect to $\sqrt{3}\times \sqrt{3}$-unit cells (Fig.~\ref{fig:3-site uc}).
On the other hand, if next-nearest neighbor repulsion is included, charge ordering patterns with 4-site unit cells (Fig.~\ref{fig:4-site uc}) become energetically competitive, allowing for striped phases and kagome-type effective sublattice charge order.
In our numerical solutions of the self-consistency equations we consider various ansaetze in 3- and 4-site unit cells, and compare their respective total energies per site to determine the global ground state.

For simplicity, we restrict our analysis to the half plane of $\mu>0$ so that in our convention, the filling factor (mean number of particles per site) $\bar{n} \geq 1$ (i.e.~hole doped scenario).
While the triangular lattice is not particle-hole symmetric, and thus the location of phases and phase boundaries may change, it is expected that for each generalized Wigner crystal at filling $\bar{n}$ there will exist a ``conjugate'' phase at filling $2-\bar{n}$.

We solve the mean-field self-consistency equations using an iterative procedure.
We work on discretized momentum-space grids with $30\times30$ unit cells for both the 3-site and 4-site ansaetze.
The free fermion Hamiltonian can be diagonalized easily in momentum space, with negligible finite-size effects.
To solve the rotor Hamiltonian, we truncate the local rotor Hilbert space (which is in principle unbounded) to finite dimension with 
$L_{\mathrm{min}}=-5$ to $L_{\mathrm{max}}=5$.
This truncation is justified since states with large $L$ would be suppressed by $U$, and we explicitly confirm its validity by noting that $Z$ is close to 1 when $U=0$ and $\varepsilon=0$, as expected.
We further note that this consideration also implies that the slave-rotor mean field approach works better for a relatively large $U$.

We work in units of the kinetic energy $t$, and employ a small but finite temperature $k_\mathrm{B} T/t = 0.01$ for numerical stability.

The mean field phase diagrams in the plane of $U$ and $\mu$ (in units of $t$) for short range repulsion $V^{\prime}=0$ is shown in Fig.~\ref{fig:both-pd}(a), and for longer-ranged repulsion $V^{\prime}=\frac{1}{\sqrt{3}}V$ in Fig.~\ref{fig:both-pd}(b).
Various charge crystal states are illustrated in Figs.~\ref{fig:both-pd}(c)-(h), and metallic states with charge dispersion on distinct sublattices in Figs.~\ref{fig:both-pd}(i)-(l).

Depending on $V'$, we find distinct charge ordering patterns:
For $V^{\prime}=0$, there are emergent honeycomb Wigner crystals at 4/3 and 5/3 filling [Figs.~\ref{fig:both-pd}(e) and \ref{fig:both-pd}(g)] as well as half-filling [Fig.~\ref{fig:both-pd}(c)].
These Mott insulating state are separated by metallic states conducting on an emergent honeycomb sublattice [Fig.~\ref{fig:both-pd}(j)] or parent triangular lattice [Fig.~\ref{fig:both-pd}(i)].
On the other hand, for $V^{\prime}=\frac{1}{\sqrt{3}}V$, charge orders of kagome type at 5/4 and 7/4 filling [Figs.~\ref{fig:both-pd}(d) and \ref{fig:both-pd}(h)] and stripe type at 3/2 filling [Fig.~\ref{fig:both-pd}(f)] are more favored than order with $\sqrt{3}\times\sqrt{3}$-periodicity.
These states are accompanied by metallic states dispersing on respective sublattices [Fig.~\ref{fig:both-pd}(k) and \ref{fig:both-pd}(l)].

States of commensurate fractional fillings in the phase diagrams are all insulators with $\sqrt{Z_i}=0$, and the compressibility $\partial n/\partial \mu=0$ because the charge per site $n_i$ is quantized in a range of chemical potential $\mu$.
In these parameter regimes, we use perturbation theory in $K/U$ in order to obtain finite (short-ranged) spin correlations determined by $\braket{\eu^{\iu(\theta_i-\theta_j)}}$ as introduced in Sec.~\ref{sec:insulating states}, with details discussed in the following subsections.

\begin{figure}
    \centering
    \includegraphics[width=\columnwidth]{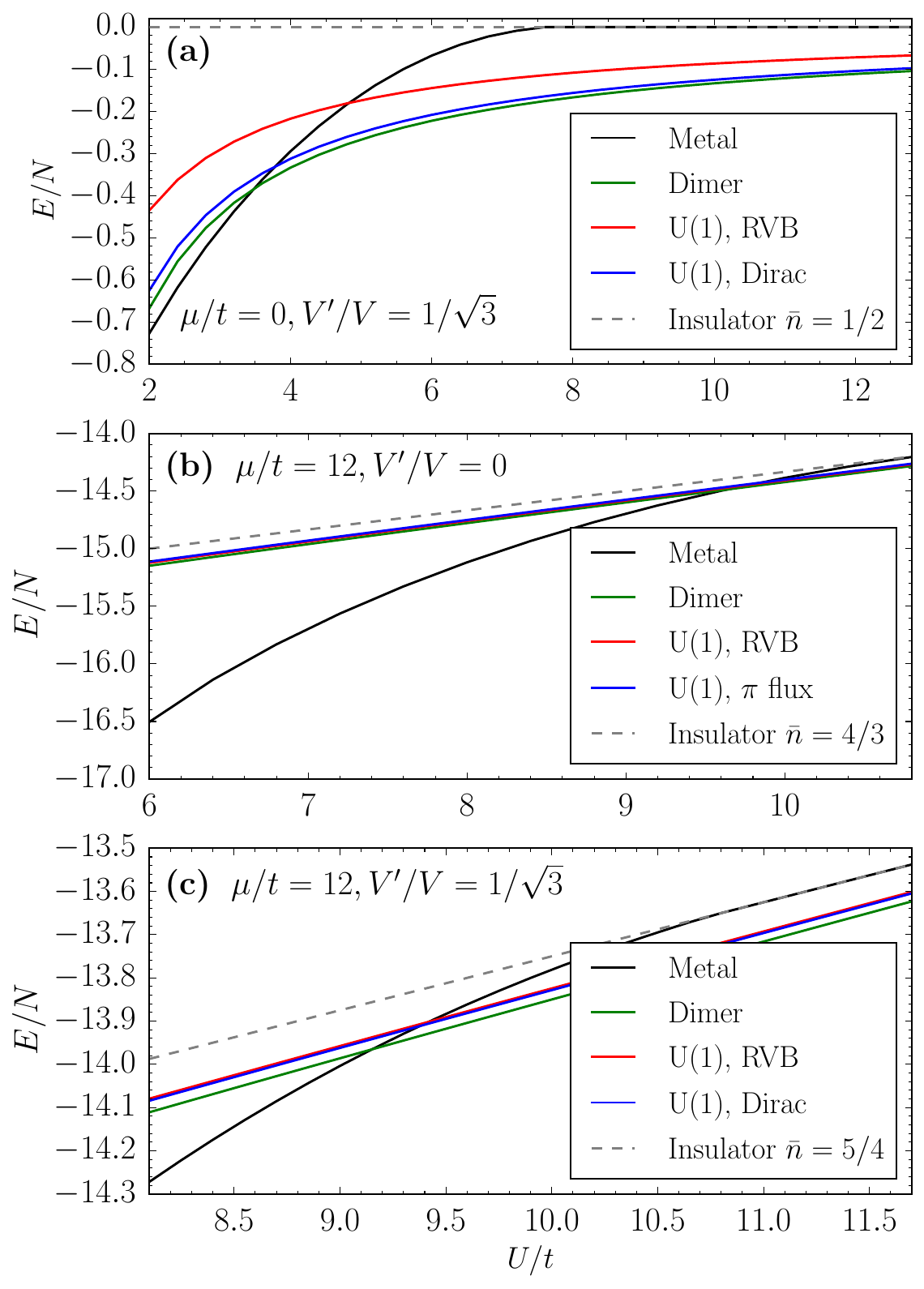}
    \caption{Energy per site $E/N$ as a function of $U/t$ for different candidate states on different Wigner crystals. (a) Triangular charge crystal (with $\bar{n}=1$), (b) Honeycomb charge crystal (with $\bar{n}=4/3$) (c) kagome charge crystal ($\bar{n}=5/4$). The dashed lines indicate the energy of ``trivial'' insulating states without any spinon hopping.}
    \label{fig:energy evolution}
\end{figure}

When some of the $\langle \eu^{\iu \theta_i} \rangle \neq 0$, there is a finite quasiparticle weight and the system is in a compressible metallic state [corresponding to incommensurate particle numbers in the phase diagram Fig.~\ref{fig:both-pd}(a) and \ref{fig:both-pd}(b)].
In addition to the metallic state corresponding to particles dispersing on the triangular lattice (with uniform $\langle \eu^{\iu \theta_i} \rangle \neq 0$), we also find states where particles disperse on a honeycomb sublattice formed by two of the three sites in the $\sqrt{3}\times \sqrt{3}$ unit cell, and kagome or stripe sublattices formed by three or two sites in the $2\times 2$ unit cell, while the remaining site is doubly occupied, leading to $\langle \eu^{\iu \theta_i} \rangle = 0$ for the corresponding $i$.
The metallic states conducting on different sublattices are sketched in Figs.~\ref{fig:both-pd}(i)-(l).
While the uniformly dispersing metallic state on the parent triangular lattice is present in both phase diagrams of $V^{\prime}=0$ and $V^{\prime}=\frac{1}{\sqrt{3}}V$, the honeycomb metal is only favored for $V^{\prime}=0$, and the kagome and stripe metals are more competitive in the latter case.

At a fixed chemical potential, upon increasing interaction strength $U$ (simultaneously also increasing $V$, $V^{\prime}$ proportionally), the metallic state enters an adjacent insulating charge-ordered state through a first-order phase transition, in contrast to a continuous Mott transition with a spectral weight going to 0, as also shown in Fig.~\ref{fig:energy evolution} in next subsection.
The critical $U$ for a continuous Mott transition is marked by the vanishing of the quasiparticle weights $Z_i$ and can be obtained by applying a perturbative approximation on $Z_i$, as shown in Ref.~\cite{florens04}.
From Eq.~\eqref{eq:Hr_metal}, and making use of Hellmann-Feynman theorem, we have
\begin{equation}
    \sqrt{Z_i}\equiv
    \braket{\eu^{\iu\theta_i}}
    =\frac{4U\sqrt{Z_i}\Tilde{K_i}}{U^2-4(UL_i+\sum_jV_{ij}L_j+h_i)^2} \label{eq:critc_U}
\end{equation}
where $\Tilde{K_i}=\sum_{j\in \mathrm{n.n}(i)}K_{ij}$ is the sum over nearest sites in the corresponding metallic sublattice (since we are keeping only nearest hoppings). At the boundary of the insulating state, $h_i=\epsilon_0=-\mu$ to satisfy the constraint Eq.~\eqref{eq:constr}. Eliminating $Z_i$ on both sides, we get an equation of the critical interaction strength $U_C$ and $V_{ijC}$.
\begin{equation}
    U_C^2-4(U_CL_i+\sum_{j}V_{ijC}L_j-\mu)^2=4\Tilde{K_i}U_C
\end{equation}

The value of $\Tilde{K_i}$ can be calculated from the self-consistency equation Eq.~\eqref{eq:Keff}, which would be independent on $\sqrt{Z_i}$ if the non-zero $\sqrt{Z_i}$ is uniform on corresponding sublattices with $h=-\mu$ near transition.

\subsection{Spin liquids and dimerized states in insulating phases}

When all $\sqrt{Z_i}=0$, all electronic quasiparticle weights vanish and thus the system enters an insulating regime.
Distinct insulating states (at identical filling/charge order) can be characterized by their corresponding spin states.
Focusing on non-magnetic states such as spin liquids and dimer states, as stated in Sec.~\ref{sec:insulating states} and Sec.~\ref{sec: classif}, the mean-field parameters $K_{ij}$ will generically be non-zero and can be used to classify various ansatz states corresponding to their respective invariant PSG.

In the following, we analyze self-consistent solutions to the mean-field equations corresponding to symmetry-allowed spin liquid states, as well as dimerized solutions. 
While we find that for all insulating states the dimerized state always has the lowest energy in our mean-field calculation (with perturbative corrections), various spin liquid states are competitive in energy with respect to the mean-field Hamiltonian.

\subsubsection{Triangular charge crystal (half-filling)}
\begin{figure}
  \centering
  \subfloat[RVB state $|K|/t=0.330$]{\includegraphics[width=0.33\linewidth]{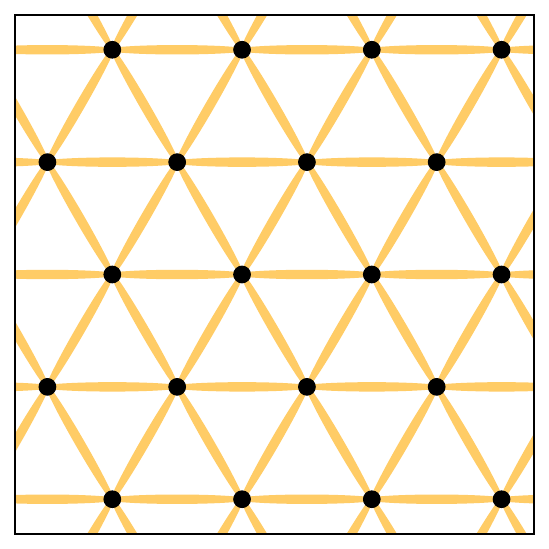}\label{fig:bond trig RVB}}
  \subfloat[Dirac state $|K|/t=0.395$]{\includegraphics[width=0.33\linewidth]{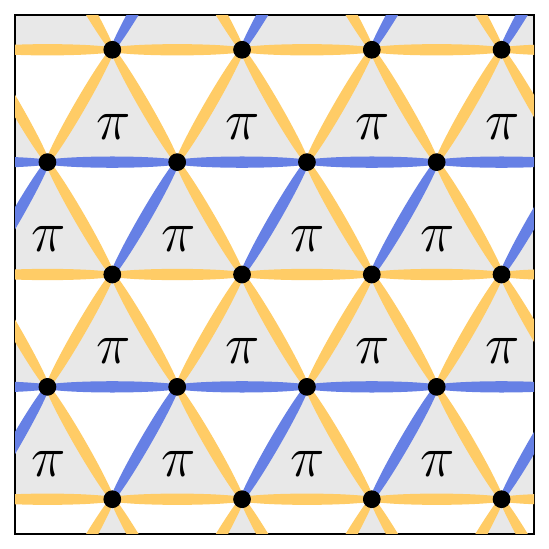}\label{fig:bond trig Dirac}}
  \subfloat[Dimer state $|K|=1$ or 0]{\includegraphics[width=0.33\linewidth]{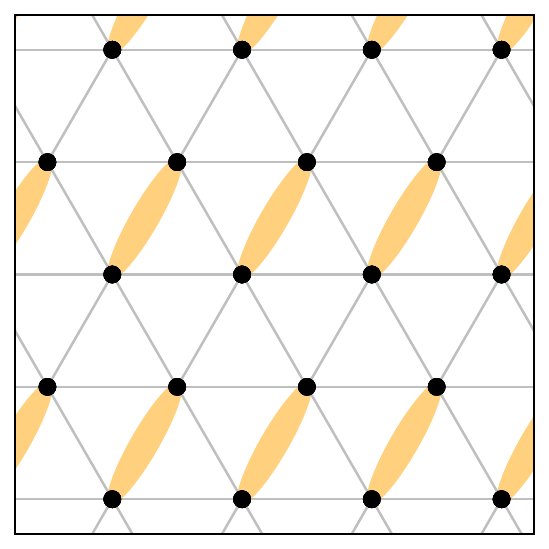}\label{fig:bond trig dimer}}
  \caption{Visualization of spin liquid ansatz and dimer states on the moiré triangular lattice at half-filling $\bar{n} = 1$. Orange bonds denote positive spinon hopping amplitudes and, by self-consistency, rotor couplings $K_{ij}$, while blue bonds denote negative hoppings and couplings. The width of the bonds is in proportion to the amplitude of rotor coupling $K_{ij}=\sum_{\sigma}\braket{f^{\dagger}_{i,\sigma}f^{\vphantom\dagger}_{j,\sigma}}$ on the respective bond. In all cases, $\mu=0, U/t=10, V/U=1/4$, and $V^{\prime}/V=1/\sqrt{3}$.}
  \label{fig:bond triangular}
\end{figure}
At half-filling, every site of the moiré triangular lattice is occupied by one charge ($L_i=0$).
While charges are localized to sites, spinons can hop on the triangular lattice with nonzero $K_{ij}$, and give rise to spin liquids and dimerized states.

\begin{figure}[t]
    \centering
    \includegraphics[width=\linewidth]{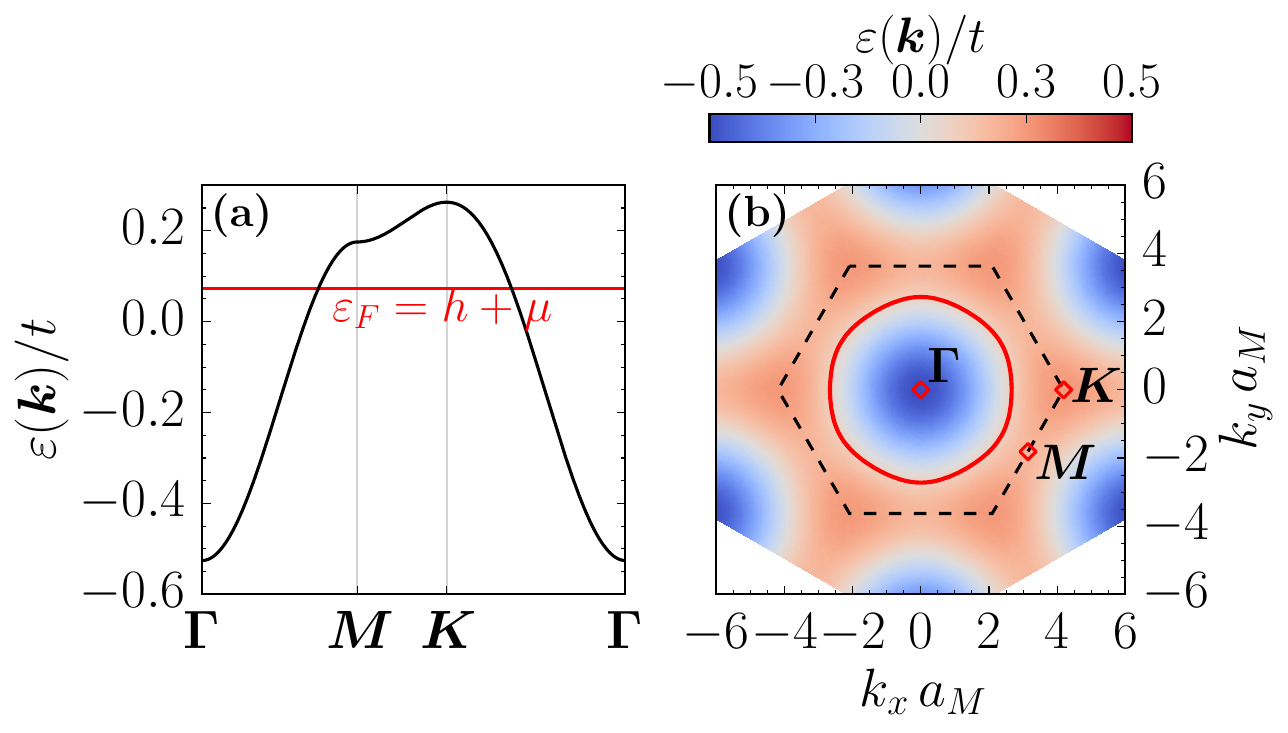}
    \caption{Spinon band structure of the 0-flux RVB state on the moiré triangular lattice, for $\mu=0$, $U/t=10$, $V/U=1/4$, and $V^{\prime}/V=1/\sqrt{3}$. The hopping is renormalized by the rotor coupling. (a) Cut along high-symmetry lines. (b) Plot of spinon band in the two-dimensional (hexagonal, dashed) Brillouin zone, with the red circle denotes the spinon fermi surface. $a_M$ denotes the lattice constant of the triangular moir{\'e} superlattice.}
    \label{fig:trig RVB band}
\end{figure}

\begin{figure}[t]
    \centering
    \includegraphics[width=\linewidth]{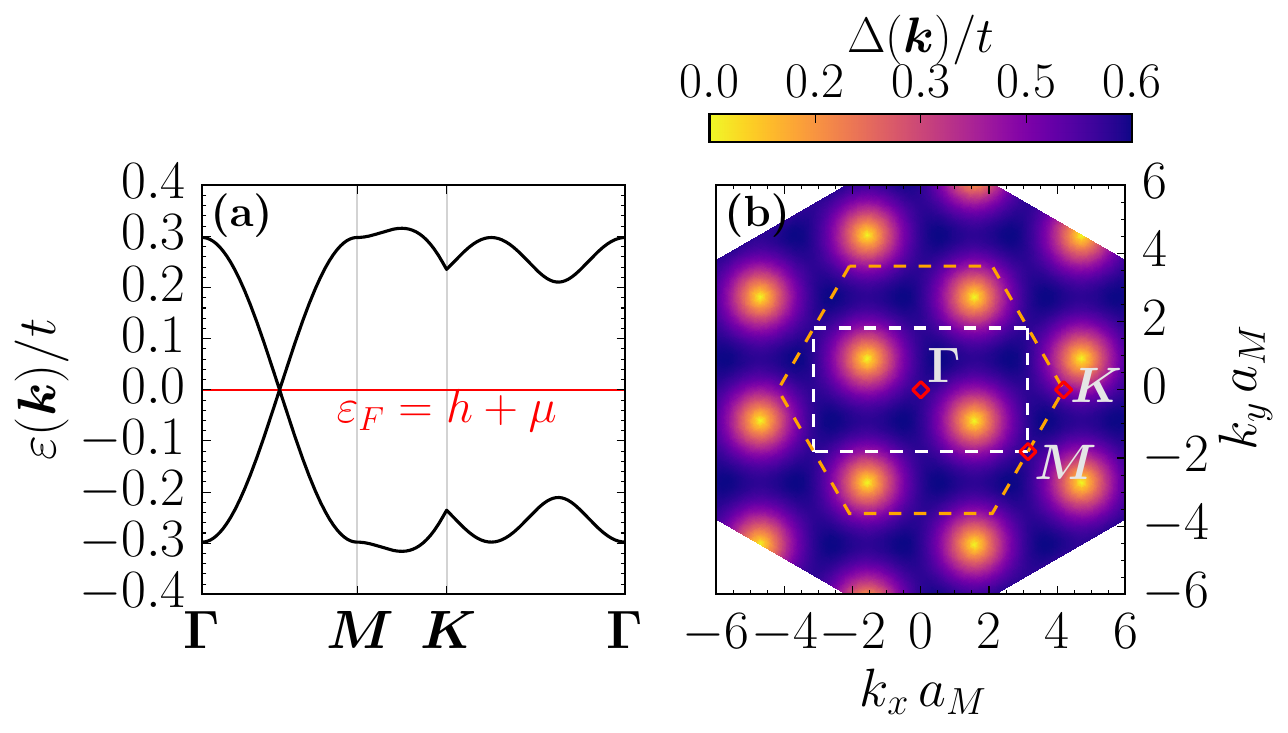}
    \caption{Spinon band structure of the Dirac state on triangular lattice (lattice constant $a_M$), for $\mu=0$, $U/t=10$, $V/U=1/4$, and $V^{\prime}/V=1/\sqrt{3}$. (a) Cut along high-symmetry lines. (b) Energy difference $\Delta(\bvec{k})/t$ between highest filled and lowest unoccupied band as a function of momentum. The orange hexagon denotes the Brillouin zone of the original triangular lattice, and the white rectangle denotes the Brillouin zone after doubling the unit cell.}
    \label{fig:trig Dirac band}
\end{figure}

Assuming that all microscopic symmetries of the system (space group, time reversal and spin rotation symmetries) are preserved, all fully symmetric spin liquid states with $\Uone$ gauge group can be classified by their respective invariant PSG as introduced in Sec.~\ref{sec: classif}.
On the triangular lattice, restricting to nearest-neighbor hoppings, there are only two distinct fully symmetric spin liquid states \cite{lu16,iqbal16}: 
\begin{enumerate}[label=(\roman*)]
    \item {The RVB state with uniform real hoppings on every bond, which corresponds to a $\Uone$ spinon Fermi surface spin liquid.}
    \item{The staggered flux state, with an emergent $\pi$ flux for the spinons in every other triangular plaquette (this can be achieved by purely real hoppings with identical magnitude, but a particular sign pattern). The spinon spectrum then exhibits Dirac nodes, and the spin state corresponds to a $\Uone$ Dirac spin liquid.}
\end{enumerate}
Restricting to particular representative (gauge-fixed) ansatz states, we find that both the RVB state and staggered flux states are solutions to our self-consistency equations.
The form of the rotor coupling as well as spinon hopping parameters (which are proportional in magnitude to each other following the perturbative calculationin Sec.~\ref{sec:insulating states}) as a result from the self-consistency calculation are shown in Figs.~\ref{fig:bond trig RVB} and \ref{fig:bond trig Dirac}.
For the RVB ansatz state (i), with the gauge choice commensurate with the full translational symmetry of the triangular lattice, there is a single band and at the self-consistent point the effective spinon Fermi energy $\varepsilon_F=h+\mu$ implies that the system is half-filled.
The bandstructure is shown in Fig.~\ref{fig:trig RVB band}.
For the staggered flux state (ii), the translational symmetry is realized projectively such that any gauge-fixed configuration of hopping parameters requires two inequivalent sites and a doubling of the unit cell to accommodate the flux pattern. 
As a result, there are two bands featuring two Dirac points in the Brillouin zone for the doubled unit cell. The corresponding bandstructure is shown in Fig.~\ref{fig:trig Dirac band}. 

Allowing for the spontaneous breaking of the lattice's rotational symmetries (but still preserving the translational symmetry with respect to $\sqrt{3} \times \sqrt{3}$ unit cells), additional states can be solutions to the self-consistency equations.
For example, an effective (anisotropic) square lattice can form if $K_{ij}=0$ on all bonds along a certain direction (say, along $\hat{x}$). The spinon dispersion develops again Dirac points if there is a $\pi$-flux in each effective square plaquette.

To find the ground state, we compare the total mean-field energies of various ansatz states.
Throughout the half-filled insulating state, dimer states have the lowest energy, as shown explicitly for selected cuts through parameter space in Fig.~\ref{fig:energy evolution}.
This is in accordance with previous studies that found dimer phases as the mean-field ground states, with a projection back to physical Hilbert space and/or inclusion of gauge fluctuations capable of stabilizing spin liquid states \cite{hastings00,rokhsar90}. We leave a systematic investigation of gauge fluctuations of discussed mean-field states for future study.

The energetic competition of the metallic state, the trivial atomic insulator (without any spin correlations) as well as the insulators with spin dimer and spin liquid states is shown at chemical potential $\mu=0$ in Fig.~\ref{fig:energy evolution}. Upon increasing $U$, the system enters the dimer state as a first order phase transition, before it approaches the atomic insulator (the dashed line) at $U_C$ determined by Eq.~\eqref{eq:critc_U}, where $\sqrt{Z_i}$ would vanishes continuously.
The same also applies for honeycomb and kagome charge crystals below.
We also find that, even though the dimer state always has the lowest energy, the $\Uone$ Dirac spin liquid state (staggered flux) is energetically competitive.

\subsubsection{Honeycomb charge crystal (4/3 filling)}
\begin{figure}
  \centering
  \subfloat[RVB state $|K|/t=0.525$]{\includegraphics[width=0.33\linewidth]{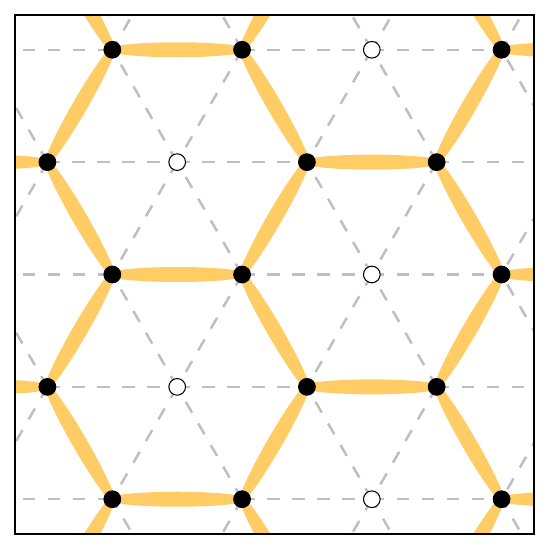}\label{fig:bond honeyc RVB}}
  \subfloat[$\pi$ flux state $|K|/t=0.502$]{\includegraphics[width=0.33\linewidth]{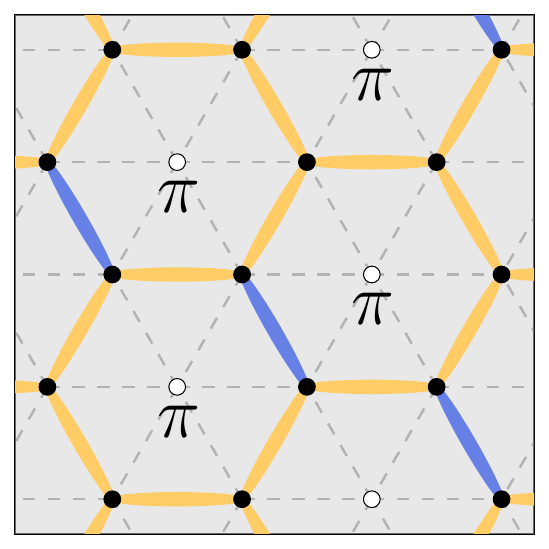}\label{fig:bond honeyc pi flux}}
  \subfloat[Dimer state $|K|/t=1$ or $0$]{\includegraphics[width=0.33\linewidth]{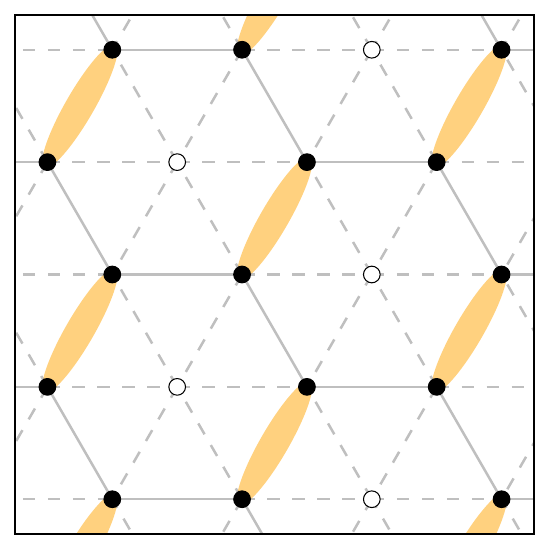}\label{fig:bond honeyc dimer}}
  \caption{Visualization of spin liquid mean-field ansatz and dimer states on the honeycomb Wigner crystal at filling $\bar{n}=4/3$. The color and width of bonds follow the same conventions and normalizations as in Fig.~\ref{fig:bond triangular}. Plaquettes which contain a $\pi$-flux are shaded grey. In all panels, $\mu/t=12, U/t=10, V/U=1/4$, and $V^{\prime}/V=0$.}
  \label{fig:bond honeycomb}
\end{figure}
For some regions in parameter space, effective honeycomb charge crystals are formed, at commensurate 4/3 filling, 
where the half-filled sites form an effective honeycomb lattice ``stuffed'' with doubly occupied sites.
While the doubly occupied sites are effectively decoupled from the system as demanded by self-consistency,
spinons can hop on the half-filled honeycomb sublattice, which possesses an effective $C_{6v}$ point group symmetry.
There are two fully symmetric $\Uone$ spin-liquid ansaetze as classified by the PSG \cite{wang10}:
\begin{enumerate}[label=(\roman*)]
    \item The RVB state, with no emergent flux.
    \item The $\pi$ flux state with a uniform $\pi$ flux in each hexagonal plaquette.
\end{enumerate}

\begin{figure}[t]
    \centering
    \includegraphics[width=\linewidth]{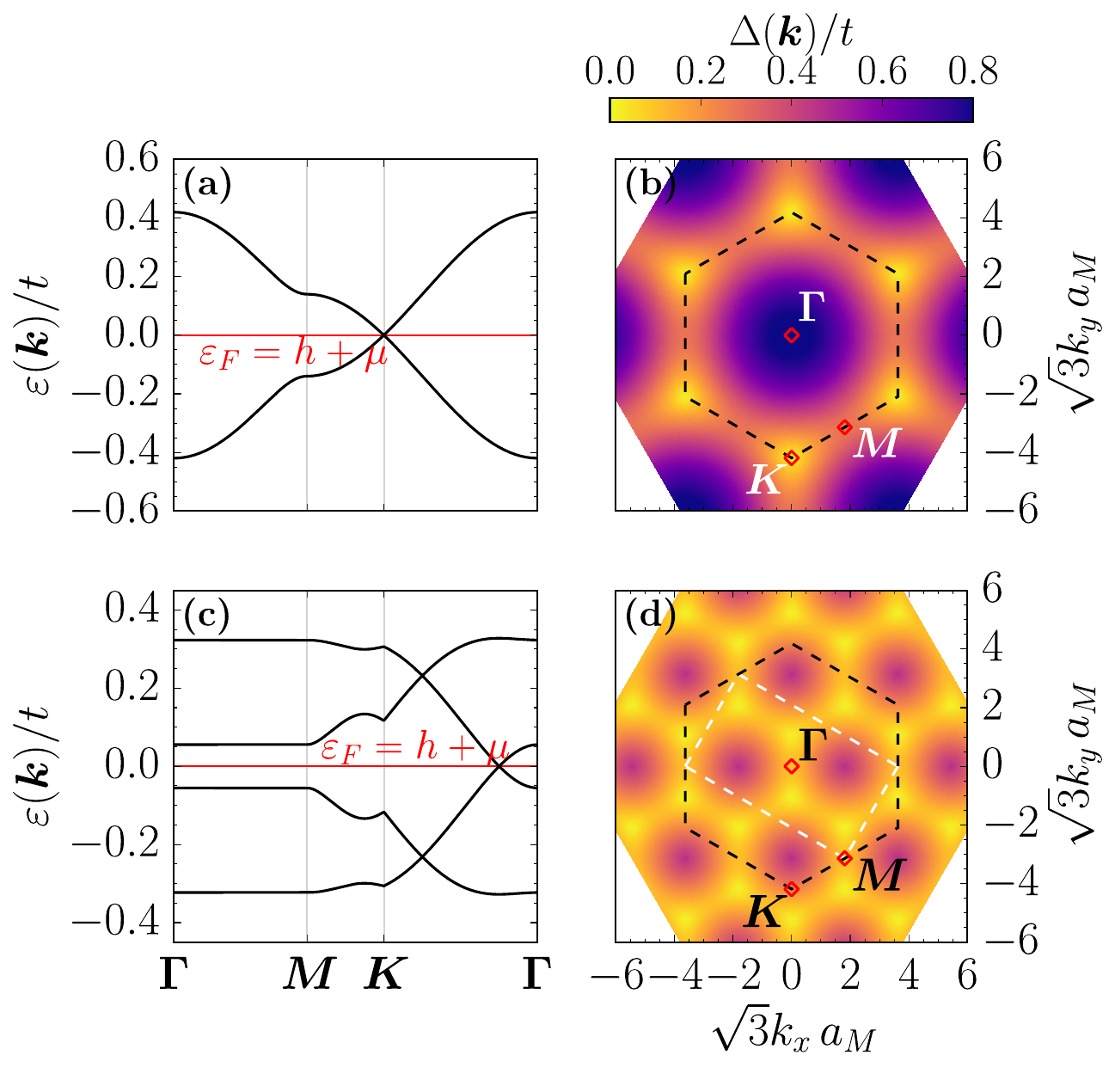}
    \caption{(a),(b) Same as Fig.~\ref{fig:trig Dirac band}, but now for the 0-flux RVB state on the honeycomb charge crystal for $\mu/t=12$, $U/t=10$, $V/U=1/4$ and $V^{\prime}=0$. The dashed hexagon in (b) denotes the Brillouin zone of the honeycomb charge crystal on the moiré triangular lattice, with lattice constant $a_M$. (c),(d) Same as (a),(b),  but now for the $\pi$-flux RVB state on the honeycomb charge crystal (lattice constant $a_M$) for $\mu/t=12$, $U/t=10$, $V/U=1/4$ and $V^{\prime}=0$. The white rectangle in (d) denotes that of the doubled unit cell with four Dirac points at the spinon Fermi level}
    \label{fig:honeyc-both-disp}
\end{figure}

We display (gauge-fixed) configurations for these two states in Fig.~\ref{fig:bond honeyc RVB} and \ref{fig:bond honeyc pi flux}.
For the 0 flux RVB state, the spinon band structure features two Dirac cones as shown in [Fig.~\ref{fig:honeyc-both-disp}(a),(b)].
For the $\pi$-flux state, the mean-field spinon Hamiltonian requires (at least) a doubling unit cell to accommodate the $\pi$ flux, and thus the four branches of bands give rise to four Dirac points in the halfed Brillouin zone [Fig.~\ref{fig:honeyc-both-disp}(c),(d)].

We show the energies of various ansatz states as a function of $U/t$ at chemical potential $\mu/t=12$ in Fig.~\ref{fig:energy evolution}.
Similar to the triangular charge crystal, we find that on the $n=4/3$ charge crystal the dimer state Fig.~\ref{fig:bond honeyc dimer} has the lowest energy of all insulating states.
We further note that the spin liquid states and the dimer state are close in energy, and the 0 flux RVB spin liquid has a consistently lower energy than the $\pi$-flux state.

\subsubsection{Kagome charge crystal (5/4 filling)}

Similar to the triangular lattice formed by charges at half-filling, the geometrical frustration of kagome charge crystal is promising for realizing quantum spin liquid states.
Again focussing on $\Uone$ spin liquids, here we consider two spin liquid states which have been found to be most competitive in energy among all PSG-allowed symmetric spin liquids on kagome lattice \cite{ran07,iqbal11}:
\begin{enumerate}[label=(\roman*)]
    \item The RVB state, with no emergent flux in either triangular and hexagonal plaquettes, denoted as [0,0].
    \item The Dirac state with zero flux on triangular plaquettes, but $\pi$ flux on hexagonal plaquettes, denoted as [0,$\pi$].
\end{enumerate}

\begin{figure}
  \centering
  \subfloat[RVB state $|K|/t=0.431$]{\includegraphics[width=0.33\linewidth]{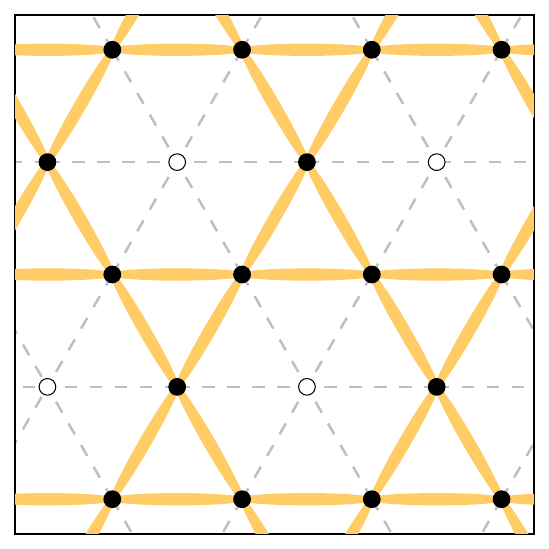}\label{fig:bond kagome RVB}}
  \subfloat[Dirac state $|K|/t=0.443$]{\includegraphics[width=0.33\linewidth]{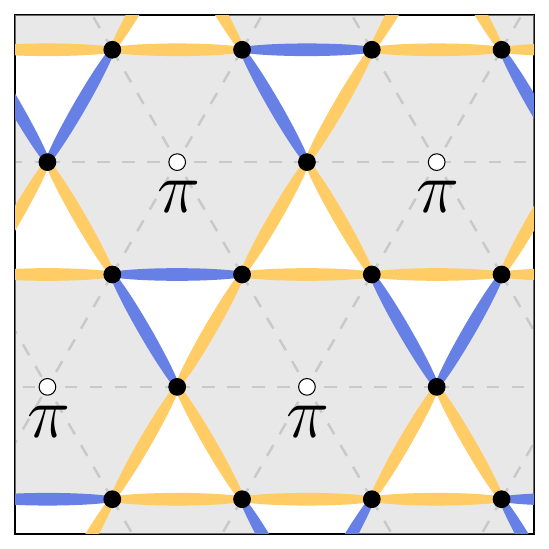}\label{fig:bond kagome Dirac}}
  \subfloat[Dimer state $|K|/t=1$ or $0$]{\includegraphics[width=0.33\linewidth]{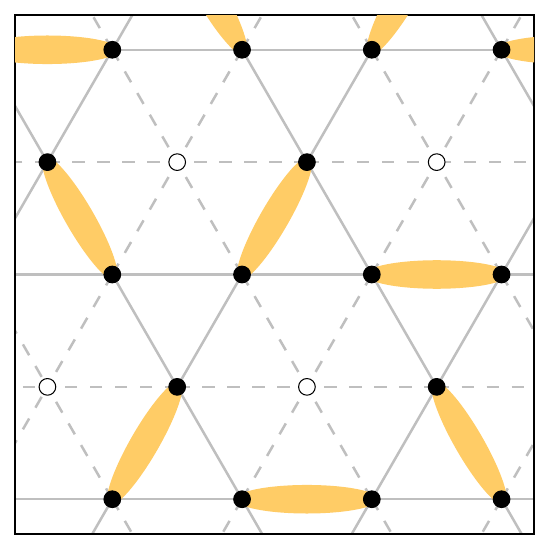}\label{fig:bond kagome dimer}}
  \caption{Visualization of mean-field spin liquid ansatz and dimer states for  kagome charge crystal at filling 5/4. The color and width of bonds have the same normalization as in Fig.~\ref{fig:bond triangular}. In all cases, $\mu/t=12, U/t=10, V/U=1/4$, and $V^{\prime}/V=1/\sqrt{3}$.}
  \label{fig:bond kagome}
\end{figure}

We illustrate the mean-field ansaetze, corresponding to particular gauge-fixed configurations, with self-consistent parameters in Figs.~\ref{fig:bond kagome RVB} and \ref{fig:bond kagome Dirac}.

\begin{figure}[t]
    \centering
    \includegraphics[width=\linewidth]{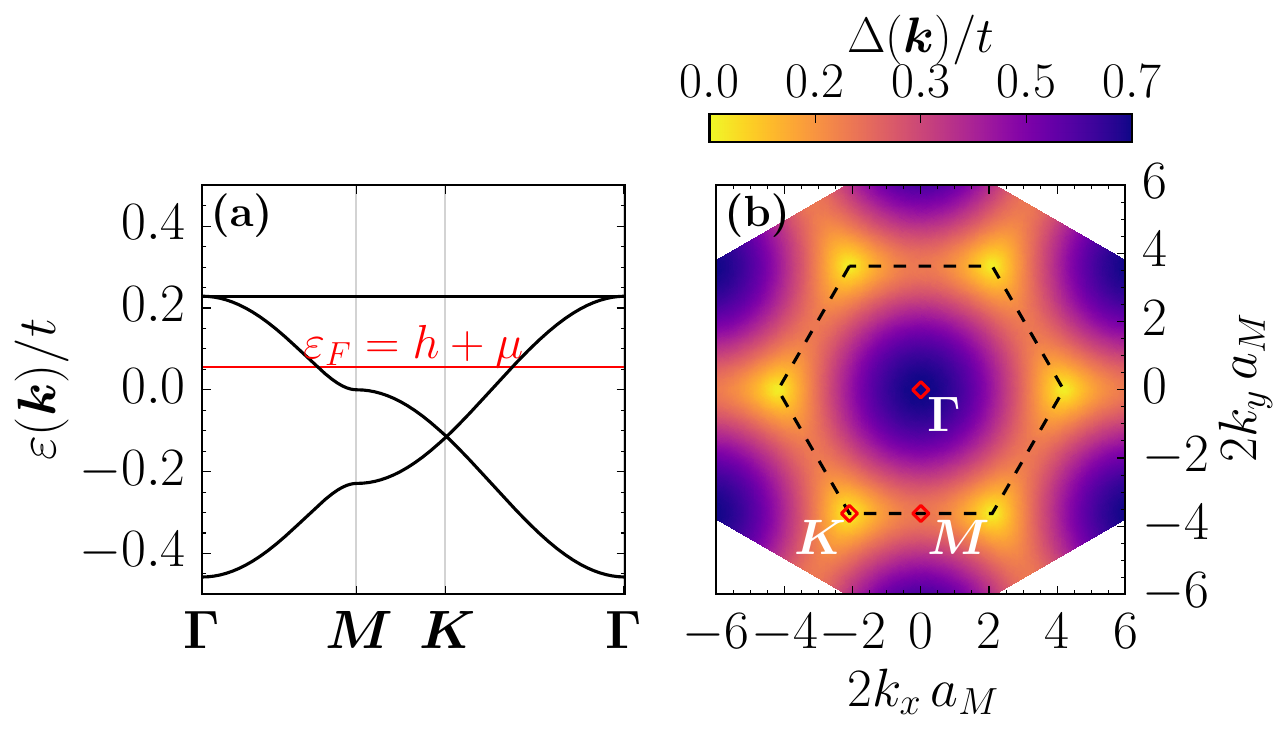}
    \caption{Same as Fig.~\ref{fig:trig Dirac band}, but now for the RVB [0,0] state on the kagome charge crystal (lattice constant $a_M$) at $\mu/t=12$,$U/t=10$,$V/U=1/4$ and $V^{\prime}/V=1/\sqrt{3}$. (a) Cut along high-symmetry path. (b) The dashed hexagon denotes the Brillouin zone of kagome charge crystal with Dirac points at $K/K^{\prime}$.}
    \label{fig:band kagome RVB}
\end{figure}

The spinon band structure of the RVB [0,0] state (corresponding to uniform hopping of spinons on the emergent kagome lattice) is shown in Fig.~\ref{fig:band kagome RVB}. 
The topmost band is found to be exactly flat corresponding to localized modes on the hexagonal plaquettes of the kagome lattice \cite{bergman08}, and there are two Dirac points at the $K/K^{\prime}$ corners of the Brillouin zone.
The spinon Fermi level $\varepsilon_F=h+\mu$ lies away from the Dirac points in order to enforce half-filling of the spinon bands, such that the RVB [0,0] state corresponds to a $\Uone$ spinon Fermi surface spin liquid.

\begin{figure}[t]
    \centering
    \includegraphics[width=\linewidth]{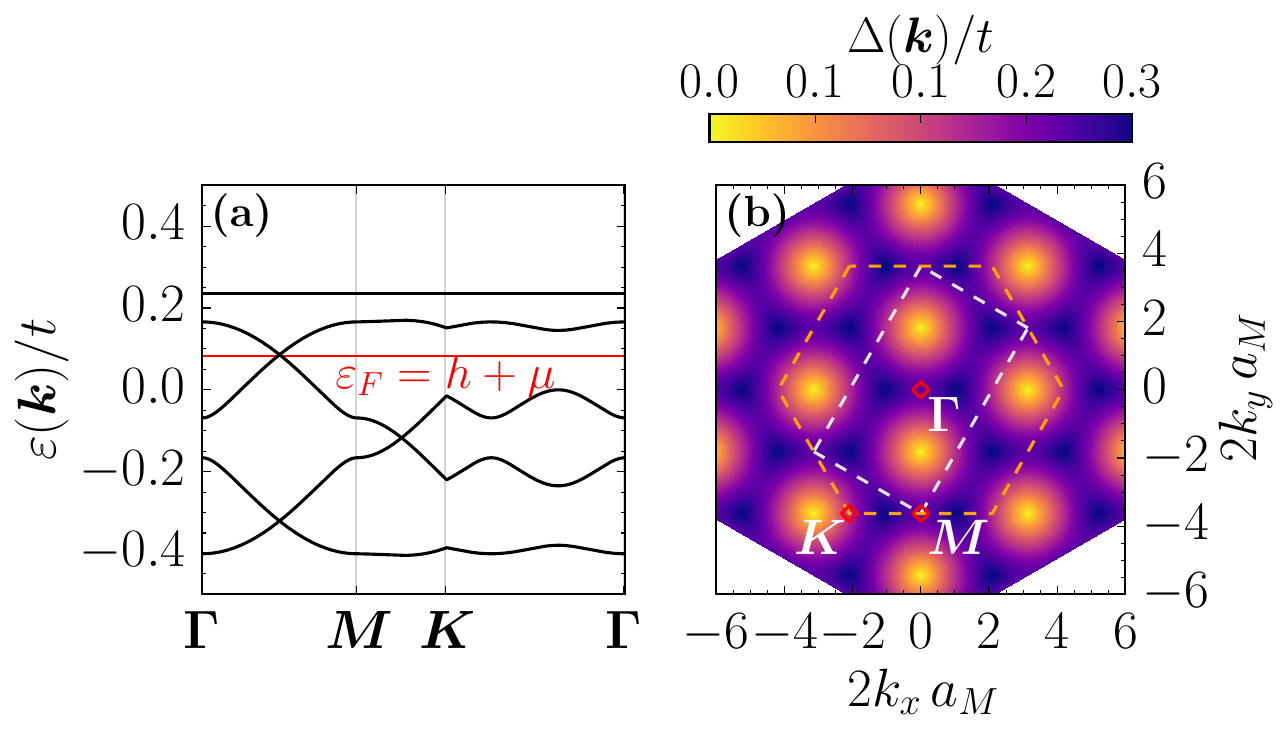}
    \caption{Same as Fig.~\ref{fig:trig Dirac band}, but now for the Dirac [0,$\pi$] state on the kagome charge crystal (lattice constant $a_M$), for $\mu/t=12$,$U/t=10$,$V/U=1/4$ and $V^{\prime}/V=1/\sqrt{3}$. (a) Cut along high-symmetry path. Note that the top flat band is twofold degenerate. (b) The dashed hexagon denotes the Brillouin zone of the kagome charge crystal, and the white rectangle denotes the Brillouin zone after doubling the unit cell, containing two two Dirac points at the Fermi level.}
    \label{fig:band kagome Dirac}
\end{figure}

We now turn to the Dirac [0,$\pi$] state, for which a doubling of unit cell is needed to accommodate a mean-field parameter configuration that contains a $\pi$ flux on hexagonal plaquettes.
The resulting band structure features a twofold degenerate flat band, as shown in Fig.~\ref{fig:band kagome Dirac}, and there are two Dirac points at the spinon Fermi level $\varepsilon_F=\mu+h$.

However, similar to the previously discussed case of the triangular lattice with unit occupancy, we find that through the phase diagram, spin dimer states (on top of the kagome charge crystal) have a lower mean-field energy than the two symmetric spin liquid ansaetze described previously.
An example of such a dimer configuration is shown in Fig.~\ref{fig:bond kagome dimer}.
The energies of different candidate states on the kagome Wigner crystal are shown in Fig.~\ref{fig:energy evolution}(c).
Note that within the insulating phase, the $\Uone$ Dirac [0,$\pi$] spin liquid is most competitive in energy to spin dimer states, while the RVB [0,0] spin liquid and trivial insulator (without any spinon hopping/hybridization) are higher in energy.

\section{Beyond mean-field theory: Stability and experimental detection} \label{Sec5}

\subsection{Spin liquids on triangular charge crystal}

Our results indicate that on the triangular lattice, the $\Uone$ Dirac spin liquid that occurs with a staggered $\pi$-flux background is the energetically most preferred spin liquid ansatz state.  This is in accord with theoretical studies of Heisenberg models valid deep in the insulating state \cite{iqbal16,PhysRevLett.123.207203}, though a U(1) spinon Fermi surface state has also been considered a strong candidate in Hubbard models \cite{PhysRevB.72.045105,PhysRevLett.95.036403}.
At low energies, the $\Uone$ Dirac spin liquid is described by quantum-electrodynamics in 2+1 dimensions (QED$_3$), where the $\SUtwo_\mathrm{pseudospin} \times \SUtwo_\mathrm{valley}$ symmetry is enhanced to an emergent $\mathrm{SU}(4)$ symmetry. QED$_3$ is believed to be stable and flow to an infrared fixed point with conformal symmetry \cite{song19,song22}.
A key consequence of this low-energy conformal field theory (CFT) is that correlation functions of order parameters close to the QED$_3$ fixed point theory are expected to fall of as anomalous power laws. For example, spin-spin correlation functions will be dominated by
\begin{equation} \label{eq:s-s-qed3}
    \langle S^\alpha(\bvec{x}) S^\beta(\bvec{y}) \rangle \sim \frac{\eu^{-\iu \bvec{K}_m \cdot (\bvec{x}-\bvec{y})}}{|\bvec{x}-\bvec{y}|^{2\Delta_\Phi}} + \hc + \dots,
\end{equation}
where $\Delta_\Phi \approx 1.02$ is the scaling dimension of the monopole operator in the QED$_3$ CFT according to latest bootstrap studies \cite{dyer13,alba22}, and $\bvec{K}_m$ denotes a wavevector at the corner of the hexagonal moiré Brillouin zone.
Here, we stress that in the context of moiré TMD discussed in this paper, the spin operators $S^\alpha(\bvec{r})$ in Eq.~\eqref{eq:s-s-qed3} refer to the $\SUtwo$ spin-valley pseudospin degree of freedom associated with a charge carrier.

We briefly comment on experimental signatures of this putative spin-liquid state in the context of moiré heterostructures.
While these systems are not amenable to conventional probes of magnetism and magnetic order such as neutron scattering due to limited sample sizes, their two-dimensional nature  allows them to be readily integrated into two-dimensional tunnel junctions.
Following experimental works on tunneling spectroscopy of magnons in van-der-Waals magnets \cite{klein18} as well as a theoretical proposal by Koenig \textit{et al.} \cite{koenig20}, we suggest that the anomalous pseudospin correlations in Eq.~\eqref{eq:s-s-qed3} of the putative Dirac spin liquid could be probed using inelastic tunneling spectroscopy, where electrons tunneling between two planar metallic substrates at the top/bottom of the heterostructure scatter off spin excitations in the moiré TMD system.
Adopting the scaling arguments of Ref.~\onlinecite{koenig20}, the inelastic contribution $I_\mathrm{inel.}$ to the tunneling current $I = I_\mathrm{el.} + I_\mathrm{inel.}$ would then exhibit an anomalous differential conductance $\du I_\mathrm{inel.} / \du V \sim V^{2\Delta_\Phi - 1}$ with $\Delta_\Phi$ as introduced in Eq.~\eqref{eq:s-s-qed3}.
As a comparison, for the Heisenberg 120$^\circ$ antiferromagnet with linearly dispersing magnons, one would expect $\du I_\mathrm{inel.} / \du V \sim V^2$.
However, we note that energy/temperature scales at which these universal power-laws are expected to be observable are unclear as of now.

\subsection{Spin liquids on honeycomb charge crystal}

We comment that $\Uone$ spin-liquid states with spinons dispersing on an effective honeycomb lattice formed by localized charges will in general be \emph{unstable}.
There are two main mechanisms for their instability:

On the one hand, we note that the gapless Dirac cone in the spinon dispersion in this state is protected by a combination of time-reversal and inversion symmetry.
As discussed in Sec.~\ref{sec:symm}, the inversion symmetry of moiré-Hubbard model on the triangular lattice occurs as an artifact of truncating the Fourier expansion of the moiré potentials $\Delta_\pm(\bvec{r})$ and $\Delta_T(\bvec{r})$ beyond leading order. Accounting for higher-order harmonics, this symmetry is broken and thus the spinon Dirac cones can be gapped out by terms of microscopic origin.
Hence, an effective field theory that describes this spin liquid state corresponds to a pure $\Uone$ lattice gauge theory in 2+1 dimensions.
However, as shown by Polyakov \cite{poly75,polyako77}, such compact $\Uone$ gauge theory is unstable against monopole proliferation.

On the other hand, even if the spinon Dirac cones are stable, in Ref.~\onlinecite{song19} it was demonstrated that $\Uone$ Dirac spin liquids on the honeycomb lattice are \emph{unstable} because there exists a monopole excitation that transforms trivially under all microscopic symmetries, and can thus proliferate.

We therefore conclude that the $\Uone$ spin liquid states for filling $\bar{n}=4/3$ are in general not expected to be stable against monopole proliferation, likely giving way to confined states such as magnetically ordered or dimer states.

\subsection{Spin liquids on the kagome charge crystal}

Amongst pure Heisenberg spin models, the kagome lattice antiferromagnet has long been known to have the strongest tendency to avoid magnetic ordering. 
While there is consensus on this point from many calculations, the precise nature of the non-magnetic ground state, and its sensitivity to weak interactions beyond nearest neighbor exchange, are under debate. 
Spontaneously dimerized valence bond solid states \cite{singh07}, as well as $\Uone$ Dirac \cite{ran07,iqbal11,song19,ran07} and $\mathbb{Z}_2$ \cite{wang06,yan2011spin,lu11,depenbrock12} spin liquid states are competitive energetically. 

We may therefore expect a spin liquid to obtain as well for the kagome charge crystal. Within our framework, we find that the most competitive spin liquid state is a $\Uone$ Dirac state, which is in accord with variational wavefunction studies for the aforementioned Heisenberg models.  The experimental signatures of a such a state would be similar to those discussed above for the triangular lattice Mott insulator.  

\subsection{Dimer states}

For all charge crystals, forming effective triangular, honeycomb, and kagome lattices, 
within our mean field theory we saw that dimerized states are actually more favorable than spin liquids and cannot be discounted.  
At the mean field level, these dimer states retain a very high degeneracy, associated to the positioning of the dimers.  This degeneracy is an artifact of the mean field treatment and is not expected to be exact.  Instead, perturbative corrections in $K/U$ beyond the mean-field approximation will induce energy differences between different dimer configurations, tending to stabilize particular dimer coverings of the lattice.  More generally, one might invoke an effective dimer model to describe the dynamics and state selection within the space of singlet dimer coverings.  Typically the ``potential'' terms within such a dimer model tend to induce an ordered configuration of singlet bonds, i.e. a valence bond crystal.  A gapped $\mathbb{Z}_2$ resonating valence bond spin liquid state is also conceivable, if the ``kinetic'' terms of such an effective dimer model dominate.  

Valence bond solid states thus appear highly plausible, and should be detectable experimentally if present.  They can occur {\em a priori} not only in the kagome charge crystal but also for other insulating fillings.  While the precise pattern of ordered dimers in such a state is hard to predict, a number of candidates have been discussed in the literature.  Rather than trying to differentiate amongst many different possible orderings, here we point out that at the grossest level they all share commonalities which suggest similar experimental signatures. In particular. any dimer pattern necessarily breaks discrete space group symmetries.  Consequently, a valence bond solid state exhibits domains and domain walls, and moreover the domains couple to lattice deformations.  These facts tend to lead to substantial {\em increases} in resistance when such symmetry breaking occurs, for example from the difficulty of propagating electrons across domain walls, against preferred directions, and from opening of gaps.  In other metals with discrete symmetry breaking these resistivity enhancements can be quite dramatic \cite{suzuki2019singular,borzi2007formation}.
In twisted TMDs, we can expect valence bond solid states to show similar resistivity enhancements tuned by gating and applied fields.  If the dimer order breaks rotational symmetry, that may also be detectable directly in transport, as it has been in nematic quantum Hall states.

\section{Summary and outlook} \label{Sec:summary}
In this paper, we have studied the triangular moir{\'e}-Hubbard model as an effective model for correlated states in moiré hetorobilayers and twisted homobilayers of transition metal dichalcogenides.
Motivated by the experimental discovery of self-organized charge lattices (generalized Wigner crystals) at various filling factors, we have employed a self-consistent slave-rotor mean-field theory to map out phase diagrams containing both insulating states as well as metallic (compressible) states as a function of interaction strength and filling, reproducing several experimentally found charge ordering patterns, where we point out that longer-ranged (next-nearest neighbor) repulsive interactions are crucial for the stabilization of kagome-type charge crystals.

By splitting electronic quasiparticles into charge rotor and fermionic spinon degrees of freedom, the slave-rotor mean-field theory (upon including charge fluctuations perturbatively) allows us to study spin states on top of the charge-ordered background, possibly including exotic spin liquid states. Such states are expected if the generalized Wigner crystals form \emph{frustrated} lattices, which occurs at half-filling (corresponding to a triangular lattice), as well as at filling $\bar{n} = 5/4$, where kagome-type lattices are stabilized.

Among all candidate ansatz states, we find that \emph{on the mean-field level}, dimer states are always energetically preferred, but certain spin liquids are found to have competitive energies, and may be realized as ground states in certain parameter reigmes after accounting for gauge fluctuations/Gutzwiller projection.
We discuss the stability of these states, as well as means of experimental detection of spin liquids and dimer states that are particularly suitable for two-dimensional heterostructures.

We note that we have restricted our study to dimer states as well as \emph{symmetric} $\Uone$ spin liquids, which are necessarily gapless.
While these states have previously been found to be prime contenders for spin-liquid ground states of frustrated Heisenberg models on the triangular and kagome lattice, $\Ztwo$ spin liquid states cannot be ruled out \emph{a priori}.
Their viability constitutes an avenue for future studies \cite{rossi23}.
Moreover, DMRG studies of the triangular lattice Hubbard model have suggested that virtual charge fluctuations at intermediate values of $t/U$ may stabilize gapped chiral spin liquids (CSL) \cite{motruk20,cookmeyer21}.
In a similar vein, one may consider vertical displacement fields which explicitly break the $\SUtwo_\mathrm{pseudospin}$ degeneracy of the moiré-Hubbard model \cite{pan20a}, leading to Dzyaloshinskii-Moriya interactions in effective spin models obtained at large $t/U$. These interactions have been suggested to induce non-collinear magnetic order as well as chiral spin liquid phases \cite{messio17,motruk22}.
A highly interesting extension of our work therefore consists in allowing for spontaneous breaking of time-reversal symmetry, which could capture possible CSL phases as well as (possibly complex) magnetic order, which we have explicitly excluded out in the manuscript at hand.

On the experimental front, we remark that detection of magnetic long-range order in moiré TMD heterostructure is still at its infancy. Recently, first evidence for ferromagnetic long-range order in R-stacked MoTe$_2$ was provided \cite{anderson23}.
While there are signatures of antiferromagnetic \emph{interactions}, for example in WSe$_2$/WS$_2$, both at half-filling \cite{tang20} as well as fractional filling \cite{tang23}, no direct detection of antiferromagnetic \emph{long-range order} in moiré TMD has been reported.
Establishing new experimental probes for magnetically ordered states in moiré TMD constitutes an important milestone for the eventual realization and detection of quantum dimer or quantum spin liquid phases in these systems.

\begin{acknowledgments}
We thank Jennifer Cano and Louk Rademaker for helpful conversations. 
UFPS is supported by the Deutsche Forschungsgemeinschaft (DFG, German Research Foundation) through a Walter Benjamin fellowship, Project ID 449890867.
ZXL is supported by the Simons Collaborations on Ultra-Quantum Matter, grant 651440 (S.S. and A.V.) from the Simons Foundation. LB and ZS were supported by the DOE, Office of Science, Basic Energy Sciences under Award No.~DE-FG02-08ER46524, and by the Simons Collaboration on Ultra-Quantum Matter, which is a grant from the Simons Foundation~(651440).
This work was performed in part at the Aspen Center for Physics, which is supported by National Science Foundation grant~PHY-2210452. 
\end{acknowledgments}

\bibliography{tmd_slaverotor_bib}
\bibliographystyle{apsrev4-2}

\clearpage

\appendix

\section{Accidental symmetry of moiré-Hubbard models} \label{sec:acc-symm}

Here, we show that the effective Hamiltonian of Ref. \cite{pan20a} for the twisted homobilayer exhibits an accidental symmetry.
Moiré bands and wavefunctions are obtained as solutions to the Schr\"{o}dinger equation $\mathcal{H} \psi = E \psi$ where
\begin{equation}
    \mathcal{H} = \left(\begin{smallmatrix} -\frac{1}{2m} \left(\hat{\bvec{k}} - \bvec{\kappa}_+ \right)^2 + \Delta_+(\bvec{r}) & \Delta_T(\bvec{r}) \\ \Delta_T^\dagger(\bvec{r}) & -\frac{1}{2m} \left(\hat{\bvec{k}} - \bvec{\kappa}_- \right)^2 +\Delta_-(\bvec{r}) \end{smallmatrix}\right),
\end{equation}
where $\hat{\bvec{\kappa}} = - \iu \bvec{\nabla}$ is the momentum operator acting in real space, and $\Delta_\pm(\bvec{r})$ and $\Delta_T(\bvec{r})$ are moiré potentials in each layer and the interlayer tunneling, respectively.
If the Bloch ansatz $\psi = \eu^{\iu \bvec{k}\cdot\bvec{r}} u_{\bvec{k}}(\bvec{r})$ solves the Schr\"{o}dinger equation with energy $E_{\bvec{k}}$, the Bloch wavefunction $w_{\bvec{k}'}(x,y) \equiv (u_{\bvec{k}'}(x,-y))^\ast$ with $\bvec{k}' = (-k_x,k_y)$ solves
\begin{multline} \label{eq:bloch-w}
    \left(\begin{smallmatrix} -\frac{1}{2m} \left(\hat{\bvec{k}} + \bvec{k} +\bvec{\kappa}_- \right)^2 + \Delta_+(x,-y) & \Delta_T^\dagger(x,-y) \\ \Delta_T(x,-y) & -\frac{1}{2m} \left(\hat{\bvec{k}} + \bvec{k}  + \bvec{\kappa}_+ \right)^2 +\Delta_-(x,-y) \end{smallmatrix}\right)w_{\bvec{k}'}
    \\= E_{\bvec{k}'} w_{\bvec{k}'},
\end{multline}
where we use that $\Delta_\pm(\bvec{r})$ is real.
Since $\bvec{\kappa}_+ + \bvec{\kappa}_-$ is equal to a reciprocal lattice vector, we can write $w_{\bvec{k}'} = \tilde{w}_{\bvec{k}'} \eu^{-\iu \left(\bvec{\kappa}_+ + \bvec{\kappa}_-\right) \cdot \bvec{r}}$. 

Now, \emph{if} we assume that $\Delta_\pm(x,-y) = \Delta_\pm(x,y)$ and $\Delta_T(x,-y) = \Delta_T^\dagger(x,y)$, the resulting Bloch problem for $\tilde{w}_{\bvec{k}'} $, with energy $E_{\bvec{k}'}$ takes the same form as $u_{\bvec{k}}$ with eigenvalue $E_{\bvec{k}}$.
Hence, there exists a combination of time-reversal and reflection symmetry that relates
\begin{equation}
    u_{\bvec{k}'}(x,y) = \eu^{\iu (\bvec{\kappa}_1 + \bvec{\kappa}_2 )} u^\ast_{\bvec{k}}(x,-y).
\end{equation}
Crucially, this symmetry requires $\Delta_\pm(x,-y) = \Delta_\pm(x,y)$ and $\Delta_T(x,-y) = \Delta_T^\dagger(x,y)$, which is satisfied by general moiré potentials only if their Fourier coefficients are restricted to the first shell of reciprocal lattice vectors.

In the case of heterobilayers with strong band offsets \cite{wu18}, the moiré-Bloch spectrum is derived by solving the Schroedinger equation for a valence band hole, $\mathcal{H} = \kappa{\bvec{\kappa}}^2 / (2m) + V(\bvec{x})$, where $V(\bvec{x})$ is the moiré potential which can be extracted from DFT simulations. Keeping the $C_3$ symmetry manifest and expanding $V(\bvec{x})$ to include only the first harmonics (i.e.~reciprocal lattice vectors only in the first moiré-Brillouin zone) leads to a spurious reflection symmetry, which is broken explicitly upon keeping higher-order terms in the expansion of $V(\bvec{x})$.


\section{Longer-ranged repulsive interactions and Wigner crystallization} \label{app:long-range-coulomb}

While the (screened) Coulomb repulsion in moiré heterostructures is generally long-ranged, it is easier for theoretical analysis and qualitative insights to truncate repulsive interactions between $n$-th nearest neighbors on the effective moiré triangular lattice.
However, we point out that such truncated models might not exhibit more complex charge-ordered (Wigner lattice) as stable thermodynamic ground states.
As a particular example, we consider the moiré-Hubbard model of Eq.~\eqref{eq:hubbard}.
At $t=0$, charges do not possess any dynamics and for filling $n \geq 1$, we can map the Hamiltonian onto an effective Ising model with $s_i = 2n_i - 3$ such that $s_i = + 1$($-1$) corresponds to doubly occupied (singly occupied) sites (a similar mapping can be performed for fillings $n < 1$).
Then, the effective Hamiltonian $H=H_U + H_\mu$ with $H_\mu = -\mu \sum_i n_i$ can be written as
\begin{equation}
    H = J \sum_{\langle ij \rangle} s_i s_j - h \sum_i s_i,
\end{equation}
which corresponds to an antiferromagnetic Ising model on the triangular lattice with nearest-neighbor coupling $J=V/4>0$ and a longitudinal field $h=(U+6 V +\mu)/2$.

As is well known, the antiferromagnetic Ising model possesses an extensive degeneracy at zero field $h=0$, which is lifted by any infinitesimal $|h| > 0$ in favor of a state with $m^z = \pm 1/3$ magnetization, which corresponds precisely to the $\sqrt{3} \times \sqrt{3}$ Wigner crystal states forming an effective honeycomb lattice with singly occupied/doubly occupied states.
Further increasing $h$, at $|h|=6J$ saturation is achieved, with all Ising spins pointing up/down (corresponding to the trivial band insulator of doubly occupied sites, or the half-filled triangular lattice).

Importantly, this argument implies that the kagome charge crystal with filling $\bar{n}=5/4$ does not exist in the grand-canonical ensemble if only nearest-neighbor repulsion is taken into account, and such that finite $V' \neq 0$ is required to stabilize the Wigner crystal states.
 
\end{document}